\documentclass[12pt]{article}
\pdfoutput=1 
\usepackage{jcappub}
\usepackage{feynmp}	
\usepackage{subfig}
\usepackage{braket}
\usepackage{cancel}
\usepackage{mathtools}
\title{On the Expected Backreaction During Preheating }
\author{C. Armendariz-Picon}
\affiliation{Physics Department\\St.~Lawrence University, \\23 Romoda Dr., Canton, NY 13617, USA.}                                            
\emailAdd{carmendarizpicon@stlawu.edu}
\makeatletter
\def\@fpheader{\relax} 
\makeatother

\abstract{In previous work we argued that the correct procedure to predict the gravitational wave signal from preheating after inflation rests on the $in$-$in$ formalism. We extend here our previous analysis to include the backreaction of the produced matter  on the motion of the inflaton and the expansion of the universe, and study how the latter affect the spectrum of the resulting gravitational waves.  The addition of backreaction demands the appropriate renormalization of divergent expectation values, which we regularize by preserving diffeomorphism invariance in a manner that is amenable to numerical integration. The very same  calculation also allows us to determine for which strength of the inflaton to matter couplings reheating is successful. We illustrate our results with the scalar version of the Starobinsky  model of inflation, and observe that it is hard to reheat the universe while keeping radiative corrections under control. }

\begin{document}
\begin{fmffile}{graphs}

 \vspace{-5cm}
 
\maketitle

\section{Introduction}
The  recent direct detection  of gravitational waves  \cite{Abbott:2016blz} has opened a new window on the universe, calling  for improved estimates of the various gravitational wave source signals.  Existing and forecasted  advances in detector technology could shed light on processes beyond the reach of other probes and may thus provide access to cosmological epochs that have mostly remained the subject of  speculation.

One of these epochs  is ``preheating" \cite{Kofman:1997yn,Allahverdi:2010xz,Amin:2014eta}.  After the end of inflation, the inflaton  oscillates around the minimum of its potential. Because of the couplings between matter and the oscillating inflaton, matter mode functions undergo  parametric resonance, in what  can be thought of as  the explosive production of matter particles. The latter  act themselves as sources of gravitational waves, which  essentially  propagate  freely  for billions of years until they are eventually detected today. 

Yet the majority of  estimates of gravitational wave production rely on  simulations \cite{Felder:2000hq,Frolov:2008hy,Sainio:2009hm,Easther:2010qz,Huang:2011gf,Child:2013ria,Lozanov:2019jff} that,  from the author's point of view, lack   proper and  rigorous justification: The heuristic claim is    that the large effective occupation numbers that result from parametric resonance practically amount to the production of classical matter waves \cite{Khlebnikov:1997di}.  These classical waves are then coupled to gravity and the resulting classical equations of motion are solved numerically, along with the linear equations for gravitational waves and their spectrum.  

But, in fact,  the matter fields are far from ``classical." Their expectation values remain zero in the quadratic theory, no matter how effective parametric resonance is. Even in the interacting theory their expectation  must vanish   if the theory is invariant under an unbroken $\mathbb{Z}_2$ transformation acting on the matter fields. And, in any case, the matter field expectation  has to   remain spatially constant because of the symmetry of the background.   In other words, during preheating the matter fields do not find themselves in a classical coherent-like state, but in the $in$ vacuum. 

In reference \cite{Armendariz-Picon:2019csc} the author argued that the proper way to address gravitational wave production during preheating involves the $in$-$in$ formalism \cite{Weinberg:2005vy}.  After all, one is just trying to predict the power spectrum of the  produced gravitational waves, and the same methods that apply to the computation of its primordial value ought to work during preheating too. The only difference between the two  is that, instead of a single tree-level diagram,  an estimate of  gravitational wave production during preheating requires the evaluation of  the one-loop diagrams in figure \ref{fig:FeynmanCon}. The occurrence of matter loops just reflects  that matter is in the $in$ vacuum, and also suggests that a tree-level  ``classical" approximation is not appropriate in this case.  

The analysis of reference \cite{Armendariz-Picon:2019csc} did not incorporate the backreaction of the matter fields on the evolution of the inflaton, nor the backreaction of the matter fields on the evolution of the spacetime metric. In this approximation, the author showed that the $in$-$in$ formalism and the standard numerical approaches  essentially agree  because the source of gravitational waves, the energy-momentum tensor, is quadratic in the matter fields, and  in the $in$-$in$ formalism the expectation of such bilinears is of the same order (but not exactly the same) as their average value in the simulations.  Nevertheless, the  issue is not as simple as it may seem at first, because the expectation of the relevant  matter field bilinears actually diverges and thus requires regularization and renormalization. Whether such an approximate agreement persists beyond the approximations made in \cite{Armendariz-Picon:2019csc} remains an open question, albeit one that we shall   address here only indirectly, as we are  mainly concerned  with the $in$-$in$ formalism. In this context, we shall just refer to \cite{Green:2020whw}, which  notes  that the evolution of interacting classical waves like those used in the traditional simulations leads to poles in correlation functions that are absent in the $in$-$in$ formalism. 

\begin{center}
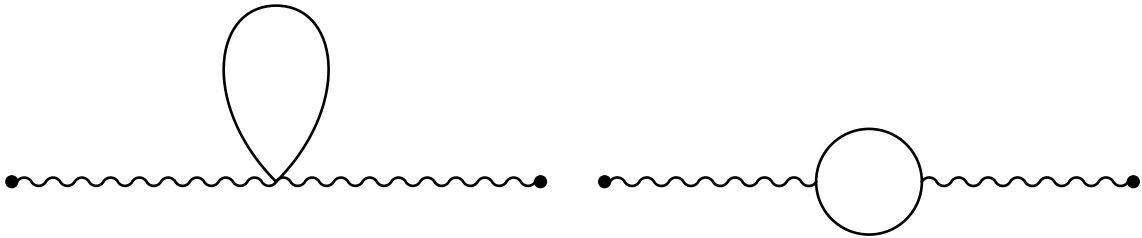
\begin{figure}
\subfloat
{
\begin{fmfgraph}(200,100) 
\fmfleft{h1} 
\fmfright{h2}
\fmfdot{h1,h2}
\fmf{wiggly}{h1,v}
\fmf{wiggly}{v,h2}
\fmf{plain}{v,v}
\end{fmfgraph}
}
\quad
\subfloat
{
\begin{fmfgraph}(200,100) 
\fmfleft{h1} 
\fmfright{h2}
\fmfdot{h1,h2}
\fmf{wiggly}{h1,v1}
\fmf{wiggly}{v2,h2}
\fmf{plain,left}{v1,v2,v1}
\end{fmfgraph}
}
\caption{\label{fig:FeynmanCon}Leading contributions to the  power spectrum of gravitational waves in the $in$-$in$ formalism. Wavy and solid lines respectively represent gravitons and matter fields.} 
\end{figure}
\end{center}

In spite of the shortcomings we just mentioned, one of the virtues of the standard numerical estimates is that they readily incorporate backreaction of the matter fields on the inflaton motion. The main goal of this work is to include backreaction into the gravitational wave analysis  of \cite{Armendariz-Picon:2019csc}, in order to bring our predictions up to par with that of the existing numerical simulations.  Our methods, however,  not only apply to the preheating stage, but   also to any other cosmological epoch. In this respect, they overlap with previous analyses of backreaction on cosmic expansion such as  those in references \cite{Calzetta:1986ey, Abramo:1997hu, Koivisto:2010pj,Markkanen:2012rh}. What sets this work apart from previous studies  is a somewhat more  careful analysis of regularization and renormalization, and the  ease to deal with cases in which analytical solutions are hard to come by and numerical methods need to be deployed.

This approach  also has applications to  the later stages of reheating, when parametric resonance ceases to be effective and the inflaton is expected to decay perturbatively. Standard analyses of the inflaton decay are based on Minkowski spacetime results, even though during reheating the universe expands, the inflaton oscillates, and, in some sense, the quantum state of the inflaton is the vacuum (as opposed to a state with non-zero occupation numbers).   Our methods bypass the concept of ``particle" by directly focusing on the renormalized energy-momentum tensor of matter, and may thus be employed to assess whether the inflaton efficiently decays into matter even in the absence of parametric resonance. In fact, our methods do not distinguish between the perturbative and non-perturbative decay of the inflaton, and can be universally applied to all of reheating and beyond. 

\section{Gravitation and Matter}
We shall consider a canonical inflationary model in which the inflaton $\phi$ minimally couples to gravity. In order for the inflaton to reheat the universe, it  is necessary for it to decay into matter and radiation at the end of inflation. We shall model both by a scalar field $\chi$ with arbitrary mass $M_0$ that we may set to zero if desired. Then, in order for the inflaton to efficiently decay, it is necessary to directly couple it to $\chi$,
\begin{equation}\label{eq:action}
\begin{split}
	S=\int d^4 x \sqrt{-g}\bigg[ \frac{M_P^2}{2}R&
	-\frac{1}{2}\partial_\mu \phi \partial^\mu \phi -V(\phi)
-\frac{1}{2}\partial_\mu \chi \partial^\mu \chi-\frac{1}{2}M_0^2\chi^2-\frac{\lambda(\phi)}{2}\chi^2\bigg].
\end{split}
\end{equation}
In this action $\lambda(\phi)$ is a ``coupling function"  that specifies the nature of the matter-inflaton couplings. Say, if we wanted to restrict our attention to renormalizable couplings we would set $\lambda(\phi)=\lambda_1\phi+\lambda_2 \phi^2$, where $\lambda_1$ and $\lambda_2$ are two couplings constants.  We focus on scalar matter because its couplings to the inflaton  typically lead to parametric resonance at the end of inflation. Heuristically, during parametric resonance the occupation numbers of the matter modes become exponentially  large, leading to the efficient production of matter particles and gravitational waves. This is not possible if matter is fermionic, because  of  Pauli's  exclusion principle \cite{Greene:1998nh}. Nevertheless, we could carry out our analysis almost verbatim for fermionic matter. The standard numerical approaches that rely on  classical solutions to the equations of motion would  presumably be inappropriate in this case, as fermion fields do not admit a classical limit. 

 We   impose for simplicity a $\mathbb{Z}_2$ symmetry of the theory under $\chi\to -\chi$, which rules out cubic couplings proportional to odd powers of $\chi$, and prevents $\chi$ from developing an expectation value, unless spontaneous symmetry breaking occurs. Actually, in the one-loop approximation we shall pursue, and provided that the expectation of the matter field $\chi$ vanishes, it shall be sufficient to consider couplings quadratic in $\chi$.    The inflaton potential $V$ is arbitrary, the only exception being that it is assumed to have a minimum that is approached as inflation ends. Without loss of generality, we can assume that  the minimum is located at $\phi=0$, and that the coupling function vanishes there, $\lambda(0)=0$.  Our intention is to consider Einstein gravity, though, as we shall see,  quantum corrections force us to introduce additional curvature terms into the action.

\section{Backreaction on the Inflaton Motion}
\label{sec:Backreaction on the Inflaton Motion}

Our next goal is to determine the impact that quantum corrections have on the motion of the inflaton. This is relevant because the inflaton couples to the matter fields $\chi$, which are in turn the sources of gravitational wave production.  More generally, we are also interested in knowing whether the inflaton actually decays, that is, whether the energy stored in its background value is efficiently dissipated into other fields.  We shall consider a homogeneous, spatially flat background spacetime,
\begin{equation}\label{eq:FRW}
ds^2=a^2(t) (-dt^2+d\vec{x}\cdot d\vec{x}),
\end{equation}
where $t$ is conformal time, and assume that the inflaton has an homogeneous background value $\bar{\phi}(t)$.  Incidentally,  in  reference \cite{Armendariz-Picon:2017llj} we argued that the inflaton does not decay \emph{during} inflation.

\subsection{Effective Equation of Motion}

Let us expand the inflaton  field $\phi$ around such  a homogeneous but otherwise arbitrary background, $\phi=\bar{\phi}+\delta{\phi}$. Expressed in terms of the fluctuations, the matter action becomes
\begin{equation}\label{eq:phi interactions}
\begin{split}
	S_\mathrm{m}=\int d^4 x \sqrt{-g} \Big[
	&\left( \Box \bar{\phi}-\bar{V}' \right)\delta\phi
	-\frac{1}{2}\partial_\mu \delta\phi \partial^\mu \delta \phi
	-\frac{1}{2}\bar{V}'' \delta\phi^2 \\
	&-\frac{1}{2}\partial_\mu\chi \partial^\mu \chi
	-\frac{M_0^2+\bar{\lambda}}{2}\chi^2-\frac{\bar\lambda'}{2}\, \delta\phi\, \chi^2-\frac{\bar\lambda''}{4}\, \delta\phi^2\, \chi^2+\cdots
	\Big].
\end{split}
\end{equation}
Here and in the following, a prime denotes a derivative with respect to $\phi$, and a bar over a function  indicates evaluation at $\bar{\phi}$.  Note that the term linear in $\delta\phi$ does not vanish because we do not assume that $\bar\phi$ satisfies its classical equation of motion. 

The presence of couplings between $\delta\phi$ and $\chi$ indicates that the inflaton influences the dynamics of the matter fields.  This is obvious from the effective mass of $\chi$, which  depends on the background value of the inflaton through the coupling function $\bar\lambda$. Diagrammatically, this influence is captured by the self-energy  diagrams in figure \ref{fig:Self Energy}.  These corrections could be inserted into any of the matter propagators in figure \ref{fig:FeynmanCon} to determine how they affect the production of gravitational waves.  We  restrict our attention to  one-loop diagrams for simplicity. 

As the inflaton oscillates around the minimum of its potential after the end of inflation,  the mode functions of the matter field $\chi$ experience parametric resonance, whereas those of the inflaton field do not. Hence, we would expect the dominant self-energy contributions to arise from the diagram with matter, rather than the inflaton, in the loop.  Alternatively, we could consider a theory in which the inflaton couples to  $\mathcal{N}$  identical matter fields and restrict our attention to the leading contribution in a large $\mathcal{N}$ expansion. In either case, the dominant matter self-energy correction stems from the tadpole diagram for the inflaton in figure \ref{fig:Self Energy}. Such  tadpoles are absent if we demand the condition
\begin{equation}\label{eq:no tadpole}
	\braket{\delta\phi}=0,
\end{equation}
that is, when the sum of all diagrams with a single external $\delta\phi$ vanishes.  Both in the $in$-$in$ and $in$-$out$ formalisms, demanding (\ref{eq:no tadpole}) is actually equivalent to the perhaps more familiar condition on the quantum effective action
$
	\delta\Gamma/\delta \bar{\phi}=0,
$
from which one typically derives the quantum-corrected equations of motion for the inflaton.  If the background is such that condition (\ref{eq:no tadpole}) is satisfied, the impact of the inflaton on the evolution of the matter fields is mainly determined by its background field value $\bar\phi$. The enforcement of equation (\ref{eq:no tadpole}) as an alternative for the evaluation of the effective potential  is known as ``Weinberg's tadpole method"  \cite{Weinberg:1973ua}.

\begin{center}
\begin{figure}
\subfloat
{
\begin{fmfgraph}(200,100) 
\fmfleft{h1} 
\fmfright{h2}
\fmftop{vt}
\fmf{plain}{h1,vb}
\fmf{plain}{vb,h2}
\fmf{dashes,tension=0}{vb,vt}
\fmf{plain}{vt,vt}
\end{fmfgraph}
}
\quad 
\subfloat
{
\begin{fmfgraph}(200,100) 
\fmfleft{h1} 
\fmfright{h2}
\fmf{plain}{h1,vb}
\fmf{plain}{vb,h2}
\fmf{dashes}{vb,vb}
\end{fmfgraph}
}
\caption{Self-energy corrections to the matter propagator at one loop. The left diagram is a tadpole diagram for the inflaton. Solid and dashes lines respectively represent the matter and inflaton fields.  \label{fig:Self Energy} } 
\end{figure}
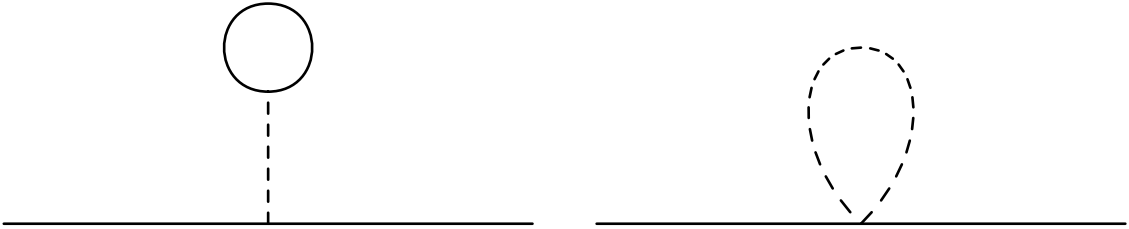
\end{center}

In order to determine  which background $\bar\phi$ fulfills equation (\ref{eq:no tadpole}), we shall simply evaluate the left hand side of equation (\ref{eq:no tadpole}) using the action in equation (\ref{eq:phi interactions}). Because  we are not assuming that $\bar\phi$ satisfies the classical field equation, there is a non-zero interaction vertex linear in   $\delta\phi$ and proportional to $\Box \bar{\phi}-\bar{V}'$.  Cancelling the $\delta\phi$ propagator we can thus cast the quantum corrected equation of motion as
\begin{subequations}\label{eq:reg eom}
\begin{equation}
	-\Box\bar\phi+\Braket{\frac{\partial V}{\partial \phi}}=0,
\end{equation}
where in the one-loop approximation
\begin{equation}\label{eq:Veff}
	\Braket{\frac{\partial V}{\partial \phi}}=\bar{V}' +\frac{\bar\lambda'}{2}\braket{\chi^2}.
\end{equation}
\end{subequations}
In this formula $\braket{\chi^2}\equiv \braket{\mathrm{in}|\chi^2|\mathrm{in}}$ is the expectation value of $\chi^2$, where the $\chi$'s are  free fields. Clearly, the interaction with $\chi$ effectively changes the inflaton potential, as one may have expected. Because they are subdominant during preheating, and in order to keep our presentation unencumbered, we are neglecting inflaton self-interactions. The latter could be easily included in our analysis nevertheless. At one loop they would give an extra contribution $(1/2)\bar{V}'''\braket{\delta\phi^2}$ to the right-hand side of equation (\ref{eq:Veff}).
Note that we can identify the $in$-$in$ expectation of the field $\phi$ with its background value, $\braket{\mathrm{in}|\phi|\mathrm{in}}=\bar\phi$, because of condition (\ref{eq:no tadpole}). We shall refer to $\braket{\partial V/\partial \phi}$ as the ``driving term."

\subsection{Evaluation of the Driving Term}
To study the evolution of the inflaton we thus  need to evaluate the expectation of $\chi^2$. We expand the free field $\chi$  in creation and annihilation operators as usual,
\begin{equation}
	\chi=\frac{1}{\sqrt{\mathcal{V}}}\sum_{\vec{k}} (w_k e^{i\vec{k}\cdot\vec{x}} a_{\vec{k}}+ w_k^* e^{-i\vec{k}\cdot \vec{x}} a^\dag_{\vec{k}} ),
\end{equation}
where $\mathcal{V}$ is the (finite) spatial volume of the universe, which we shall take to infinity momentarily.  In the spacetime (\ref{eq:FRW}) the mode functions $w_k$ satisfy the mode equation
\begin{equation}\label{eq:mode equation}
	\ddot{w}_k+2\mathcal{H}\dot{w}_k+(k^2+a^2 \kappa_0)w_k=0,
\end{equation}
where $\kappa_0\equiv M_0^2+\bar\lambda$ is the effective (inflaton-dependent)  squared mass of the matter field $\chi$, $\mathcal{H}\equiv \dot{a}/a$ is the comoving Hubble scale, and a dot denotes a derivative with respect to conformal time.   Then, assuming that matter is in the $in$ vacuum, $a_{\vec{k}}\ket{\mathrm{in}}=0$, we find 
\begin{equation}\label{eq:unren chisq}
	\langle{\chi^2}\rangle=\frac{1}{(2\pi)^3} \int d^3k \,  |w_k|^2,
\end{equation}
where we have replaced the discrete mode sum by an integral.  Note that  $\braket{\chi^2}$ is spatially constant, as a consequence of the homogeneity of the cosmological background. 

 We shall regulate the divergent mode integral (\ref{eq:unren chisq}) with  exactly the same methods described in reference \cite{Armendariz-Picon:2019csc}, to which we refer the reader for further details. Essentially,  the regularization  entails the introduction of a set of $n$ Pauli-Villars regulator fields $\chi_r$  of Grassmann parity $\sigma_r$, with the same couplings as $\chi$, but with different masses $M_r$, where $r=1,\ldots n$. Because they  couple like the original matter field $\chi\equiv \chi_0$, their effective squared masses are
  \begin{equation}\label{eq:kappai}
	\kappa_i \equiv M_i^2+\bar \lambda,
\end{equation}
and the actual expectation that enters the effective potential in equation (\ref{eq:Veff})   is 
$
{\braket{\chi^2}\equiv \sum_{i=0}^n \braket{\chi^2_i}}.
$
At the end of the calculation we shall decouple the regulator fields by sending their masses $M_r$ to infinity, leaving a finite, renormalized theory behind. 

We can identify the potentially divergent contributions in equation (\ref{eq:unren chisq}) by first introducing a cutoff at comoving momenta $k=\Lambda$  and then expanding  the mode integrals in the number of time derivatives acting on the background. The key ingredient here is   an analogous ``adiabatic" expansion of the mode functions in the number of time derivatives. The latter  is only valid at large values of $k$  or large values of $M_r^2$. Hence, we shall only be able to analytically recover the ultraviolet behavior of the original mode integral (\ref{eq:unren chisq}), or the magnitude of $\braket{\chi_r^2}$ when the  regulators  become sufficiently heavy. Fortunately, these are the only regimes we shall need to renormalize the divergences we shall encounter.  

At zeroth-order in time derivatives we find
 \begin{subequations}\label{eq:chisq}
\begin{equation}\label{eq:chisq0}
	\langle{\chi^2}\rangle^{(0)}=\frac{1}{4(2\pi^2)}\sum_i \sigma_i
	\left[\frac{\Lambda^2}{a^2}+\frac{\kappa_i}{2}\left(1-\log x_i\right)\right],
\end{equation}
and at two derivatives we obtain
\begin{equation}\label{eq:chisq2}
	\langle{\chi^2}\rangle^{(2)}=\frac{1}{24 (2\pi^2)}\sum_i \sigma_i
	\left[
	\bar{R}\left(-\frac{5}{6}+ \frac{1}{2}\log x_i\right)
	+\frac{\mathcal{H}}{a^2}\frac{\dot\kappa_i}{\kappa_i}
	-\frac{\dot\kappa_i^2}{4a^2 \kappa_i^2}
	+\frac{\ddot{\kappa}_i}{2a^2 \kappa_i}
	\right],
\end{equation}
\end{subequations}
where we have  introduced the dimensionless ratio 
\begin{equation}\label{eq:xi}
	x_i  \equiv \frac{4\Lambda^2}{a^2 \kappa_i}
\end{equation}
and used that for the metric (\ref{eq:FRW}) the Ricci scalar is $\bar{R}=6\ddot{a}/a^3$. In both equations (\ref{eq:chisq}),   the expectation depends on $\bar\phi$ through the squared effective  masses $\kappa_i$.  Note that the  leading divergent terms with no derivatives are proportional to the square of the cutoff or the field masses, whereas  those with two derivatives are at most logarithmically divergent. On dimensional grounds,  terms with higher derivatives  remain finite as the cutoff $\Lambda$ is removed or the regulators are decoupled. Say, all terms with four derivatives remain finite  in the limit $\Lambda\to \infty$; their form ranges from $H^4/\kappa_i$ to $\ddot{\kappa}^2_i/(a^4 \kappa_i^3)$, where $H\equiv \mathcal{H}/a$ is the Hubble constant.

\subsection{Renormalization of the Driving Term}

From equations (\ref{eq:chisq}), as we send $\Lambda$ to infinity the mode integral remains finite if the regulator masses and parities obey 
\begin{equation}\label{eq:sigma cond}
\sum_i \sigma_i=0, \quad \sum_i \sigma_i M_i^2 =0.
\end{equation}
This is possible  because fermionic  fields ($\sigma_i=-1$) give loop contributions  with the opposite sign as those of bosonic fields ($\sigma_i=1$). If  conditions  (\ref{eq:sigma cond}) are satisfied the theory  is finite, but the expectation still depends on the otherwise arbitrary regulator masses. Although $\braket{\chi^2}$ still appears to  depend on the physical cutoff $\Lambda/a$ through $x_i$, this dependence cancels again because of equations (\ref{eq:sigma cond}), so we might as well replace $\Lambda/a$ in $x_i$ by any other  mass scale. 

 The dependence on the regulator fields disappears if we decouple them by sending their masses to infinity, $M_r\to \infty$. Then,  their only traces left are the divergent contributions stemming from the logarithms in equations (\ref{eq:chisq}). The latter happen to contribute to the effective equation of motion  just like the  counterterms
\begin{equation}\label{eq:counterterms 1}
	S_\mathrm{ct}\supset \int d^4x \sqrt{-g}\left[-\delta d_1\,  \lambda(\phi)-\delta d_2\, \lambda^2(\phi)-\delta\xi\,  \lambda(\phi) \,R\right],
\end{equation}
which introduce additional corrections to the effective equation of motion of the inflaton field,
\begin{subequations}
\begin{equation}\label{eq:quantum eom}
-\Box\bar\phi+\Braket{\frac{\partial V}{\partial \phi}}_\mathrm{ren}=0,
\end{equation}
where we have identified the renormalized driving term with
\begin{equation}\label{eq:dVren}
\Braket{\frac{\partial V}{\partial \phi}}_\mathrm{ren}\equiv
	\bar{V}'+
	\frac{\bar\lambda'}{2}\sum_i \braket{\chi_i^2} 
	+\delta d_1 \, \bar\lambda'
	+2\delta d_2\,  \bar\lambda \bar\lambda'
	+\delta\xi\,  \bar{\lambda}'\,\bar R.
\end{equation}
\end{subequations}
Clearly, the counterterms proportional to $\delta d_1$ and $\delta d_2$ can be also  thought of as part of the inflaton potential.  The  coefficients in  equations (\ref{eq:chisq})   that diverge logarithmically as $M_r\to \infty$ are canceled by the counterterms, provided that the latter diverge like
\begin{subequations}\label{eq:phi counterterms}
\begin{align}
	2\pi^2 \delta d_1&=
	 \frac{1}{16}\sum_r \sigma_r M_r^2 \log X_r +\delta d^f_1,
	\\
	2\pi^2\delta d_2&=
	\frac{1}{32}\sum_r\sigma_r   \log X_r+\delta d^f_2, 
	\\
	2\pi^2 \delta\xi &=-\frac{1}{96}
	\sum_r \sigma_r \log X_r+\delta\xi^f\
	\label{eq:delta xi}.
\end{align}
\end{subequations}
The yet undetermined and \emph{finite} pieces of the counterterms are denoted by the  superscript ``$f$." The log-divergent pieces force us to   usher in an arbitrary inflaton scale $\bar\phi=\mu$ and the corresponding  effective mass of matter at that scale,
\begin{equation}
\kappa_\mu\equiv M_0^2+\lambda(\mu), \quad X_r \equiv \frac{\kappa_\mu}{M_r^2}.
\end{equation}

The counterterms  (\ref{eq:counterterms 1}) share the structure of the original action (\ref{eq:action}) only for particular choices of the coupling function $\lambda$. Say,  
 if $\lambda$ is a polynomial of degree $n$, the counterterms introduce corrections to the inflaton potential of degree $2n$.  This implies that renormalizability (in our limited context) demands the bare scalar field potential also be a polynomial of degree $2n$. The finite pieces of each of the coefficients in such a polynomial are then determined by appropriate renormalization conditions, as we explore below, whereas the divergent components follow from equations (\ref{eq:phi counterterms}). If we assume that the scalar field potential is renormalizable in the traditional sense, $V$ can only be a quartic polynomial, and this restricts the coupling function to ${\lambda=\lambda_1 \phi+\lambda_2 \phi^2}$. There is no wave function renormalization at one loop,  hence the absence of a counterterm proportional to the squared gradient of the inflaton. 
 
As far as the terms with zero derivatives are concerned, our discussion parallels that of field theory in Minkowski spacetime. In particular, we have basically obtained the derivative of the Coleman-Weinberg effective potential in the theory defined by (\ref{eq:action}). Indeed, terms without time derivatives are not sensitive to the expansion of the universe, and should thus reproduce the Minkowski spacetime results. With the  unrenormalized but regularized effective potential given by the integral over Euclidean four-momenta  \cite{Coleman:1973jx}
\begin{equation}
	U_{CW}=\bar V+\frac{1}{2}\sum_i \sigma_i \int \frac{d^4 p_E}{(2\pi)^4} \, 
	\log\left(1+\frac{\kappa_i}{p_E^2}\right),
\end{equation}
it is then  easy to see that $dU_{CW}/d\bar\phi$ equals $\bar{V}'+\frac{\bar\lambda'}{2} \braket{\chi^2}^{(0)}$, in agreement with our calculation. Plugging the counterterms (\ref{eq:phi counterterms}) into equation (\ref{eq:dVren}) and taking the limit ${M_r\to\infty}$ we arrive at
\begin{equation}
	\Braket{\frac{\partial V}{\partial \phi}}^{(0)}_\mathrm{ren}=
	\bar V'+
	\frac{1}{2\pi^2}\left[
	\frac{(M_0^2+\bar\lambda)\bar\lambda' }{16}\log\frac{M_0^2+\bar\lambda}{M_0^2+\lambda(\mu)}
	+\delta d^f_1 \, \bar\lambda'
	+\left(2\delta d^f_2-\frac{1}{16}\right)\bar\lambda \bar\lambda'
	\right].
\end{equation}
Because the minimum of the inflaton potential is located at $\phi=0$,  it would be  natural to choose $\mu=0$  as renormalization point,  but we would encounter then a zero-mass singularity  when $M^2_0\to 0$. It is hence convenient to choose a non-zero $\mu$. Note that the presence of two arbitrary finite counterterms indicates that two conditions are necessary to fix the actual form of the driving term at zero derivatives. 

But there is yet another contribution to the driving term at two derivatives that is sensitive to the ultraviolet, and  forces the inflaton to couple non-minimally to gravity.  It can be read off equations (\ref{eq:dVren}), (\ref{eq:chisq2})  and (\ref{eq:delta xi}), which lead to
\begin{equation}
	\Braket{\frac{\partial V}{\partial \phi}}^{(2)}_\mathrm{ren}=\frac{\bar{R}\bar{\lambda}'}	{2\pi^2}\left[\delta\xi^f-\frac{1}{48}\log \frac{M_0^2+\bar{\lambda}}{M_0^2+\bar\lambda(\mu)}
	\right].
\end{equation}
The logarithmic dependence  can be thought of as due to the running of the coupling constant $\delta\xi$ with the inflaton field. Even if $\delta\xi$ vanishes at $\bar{\phi}=\mu$, the  running reintroduces the non-minimal coupling away from that value.

 \begin{figure}
\subfloat[Unrenormalized]
{
\includegraphics[width=7.5cm]{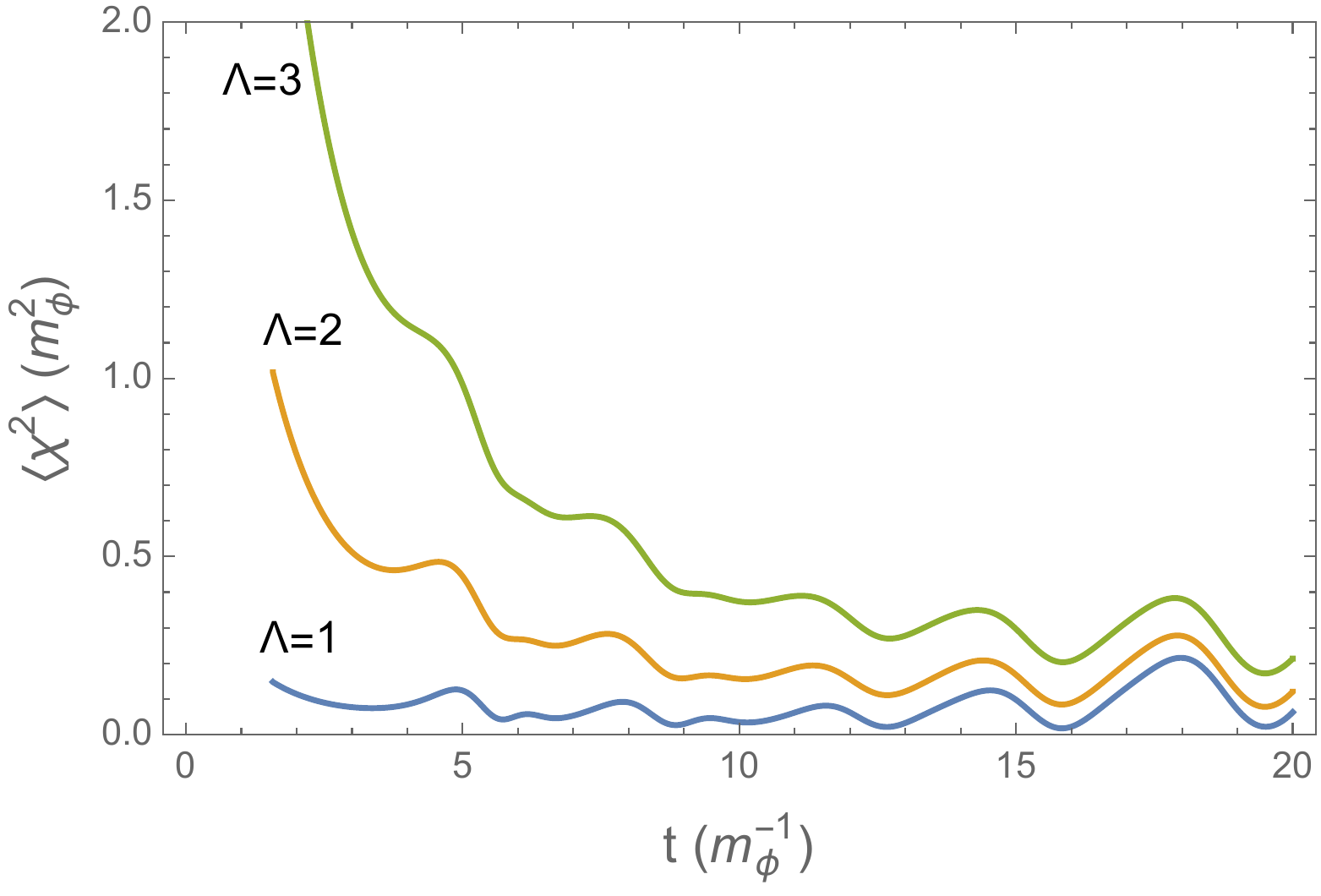}
}
\subfloat[Renormalized]
{
\includegraphics[width=7.5cm]{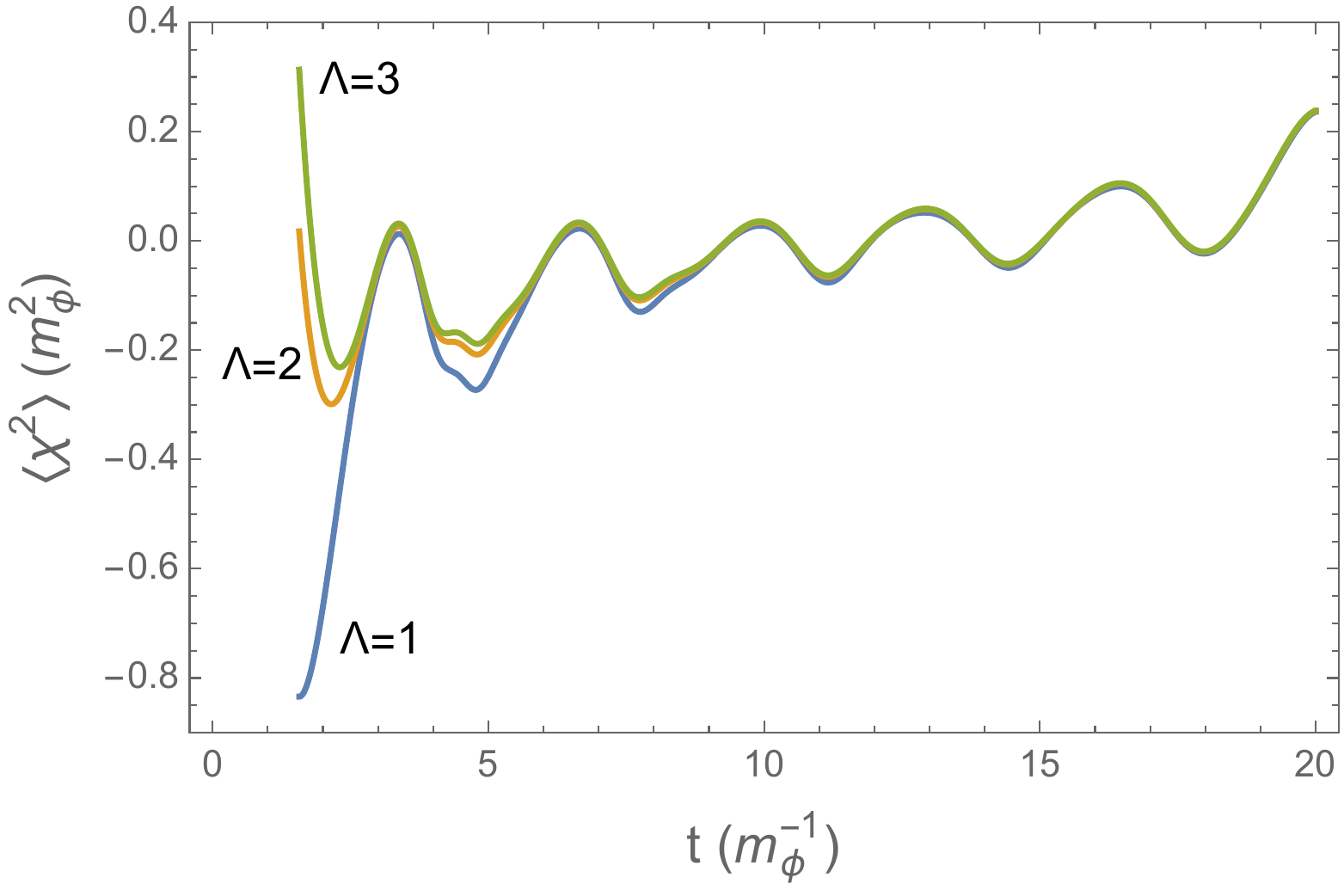}
}
\caption{Time evolution of $\braket{\chi^2}$ in the absence of backreaction for different cutoff values $\Lambda$ and fixed $q_0=10^2$.  The value of  $\Lambda$ is shown in units of $(2800/\pi^7)^{1/6}\, m_\phi$, and is chosen to   encompass all the modes that undergo parametric resonance \cite{Armendariz-Picon:2019csc}.  Panel (a) shows a clear dependence of the  unrenormalized expectation on $\Lambda$. In Panel (b), such a dependence is seen to cancel out for sufficiently large $\Lambda$.   The unrenormalized expectation is positive by construction, but note that its renormalized counterpart can become negative. \label{fig:chisqlambda} } 
\end{figure}

Yet it is also important to note that the quantum corrections that we have explicitly computed so far  are only  the first few terms in an infinite series. The renormalization conditions only affect the first two terms in such an expansion, but there is an infinite number of additional terms with higher derivatives that remain. Therefore, it is not sufficient to solely rely on the effective potential to discuss the impact of quantum corrections on the evolution of the inflaton. On dimensional grounds,   the relative size  of a correction with $2n$ time derivatives is expected to be of order $(H^{2}/\kappa_0)^n$, and is thus  likely to be negligible when $H^2\ll \kappa_0$.  In order to go beyond such  limit, it suffices to combine equations (\ref{eq:chisq}), (\ref{eq:dVren})   and (\ref{eq:phi counterterms}), which yield the  ``exact" renormalized driving term at one loop, valid at all orders in the derivative expansion,
\begin{eqnarray}\label{eq:ren driving}
	\Braket{\frac{\partial V}{\partial \phi}}_\mathrm{ren}
	\!\!\! &= &\bar V'
	+\frac{1}{2\pi^2}\Bigg\{
	\delta d^f_1\, \bar\lambda'
	+\left(2\delta d^f_2-\frac{1}{16}\right)\bar\lambda \bar\lambda' 
	+\delta \xi^f \bar\lambda' \, \bar R
	\\
	&+&\frac{\bar\lambda'}{2}\Bigg[ \lim_{\Lambda\to \infty}\int_0^\Lambda \!\!\! dk \, k^2|w_k|^2-\frac{1}{4}\frac{\Lambda^2}{a^2}-\frac{\kappa_0}{8}\left(1-\log x_\mu\right)
	-\frac{\bar R}{24}\left(-\frac{5}{6}+\frac{1}{2}\log x_\mu \right)\Bigg]
	\Bigg\},
	\nonumber
\end{eqnarray}
where we have introduced the dimensionless ratio
\begin{equation}
x_\mu\equiv \frac{4\Lambda^2}{a^2 \kappa_\mu}.
\end{equation}
One advantage of our approach is that  we are able to carry out the renormalization program while preserving diffeomorphism invariance and keeping the resulting expressions in a form suitable  for numerical integration. Though diffeomorphism invariance has not played much a role so far, it is crucial in the analysis of the backreaction on the metric. 

\subsection{Numerical Results}
\label{sec:BRIMNR}

We have implemented the renormalization of the driving term numerically, by essentially following  the same approach of reference \cite{Armendariz-Picon:2019csc}. The complete set of equations solved by our code is presented in appendix \ref{sec:Equations of Motion}. In this context, it is convenient to work with dimensionless quantities in a specific model. For illustration we choose the arguably simplest  potential and coupling functions (\ref{eq:illustration}), which in the original variables read
\begin{equation}\label{eq:example}
	V=\frac{1}{2} m_\phi^2\phi^2, \quad \lambda=\lambda_2\, \phi^2,
\end{equation}
and also happens to agree with the model most often discussed in the literature.  A welcome feature of such a $\lambda$  is that the corrections to the driving term vanish at $\bar\phi=0$. Thus, $\bar{\phi}=0$ remains a stable equilibrium point of the inflaton, even with quantum corrections taken into account. In order to avoid  the introduction of additional dimensional constants in our numerical analysis we choose
 \begin{equation} \label{eq:masses}
 	M_0\equiv 0, \quad \mu\equiv M_P.
 \end{equation}
 In particular, we assume that the matter the inflaton decays into is massless (radiation). The renormalization point $\mu=M_P$ determines the structure of the driving term at $\bar{\phi}=\mu$. It somewhat simplifies  for the values that we employ in our numerical analysis, namely, 
\begin{equation}\label{eq:finite CT 1}
	\delta d^f_1=\frac{M_0^2}{32},\quad
	\delta d^f_2=\frac{3}{64}, \quad
	\delta \xi^f=-\frac{1}{96}.
\end{equation}

\begin{figure}
\subfloat[$q_0$ dependence]
{
\includegraphics[width=7.5cm]{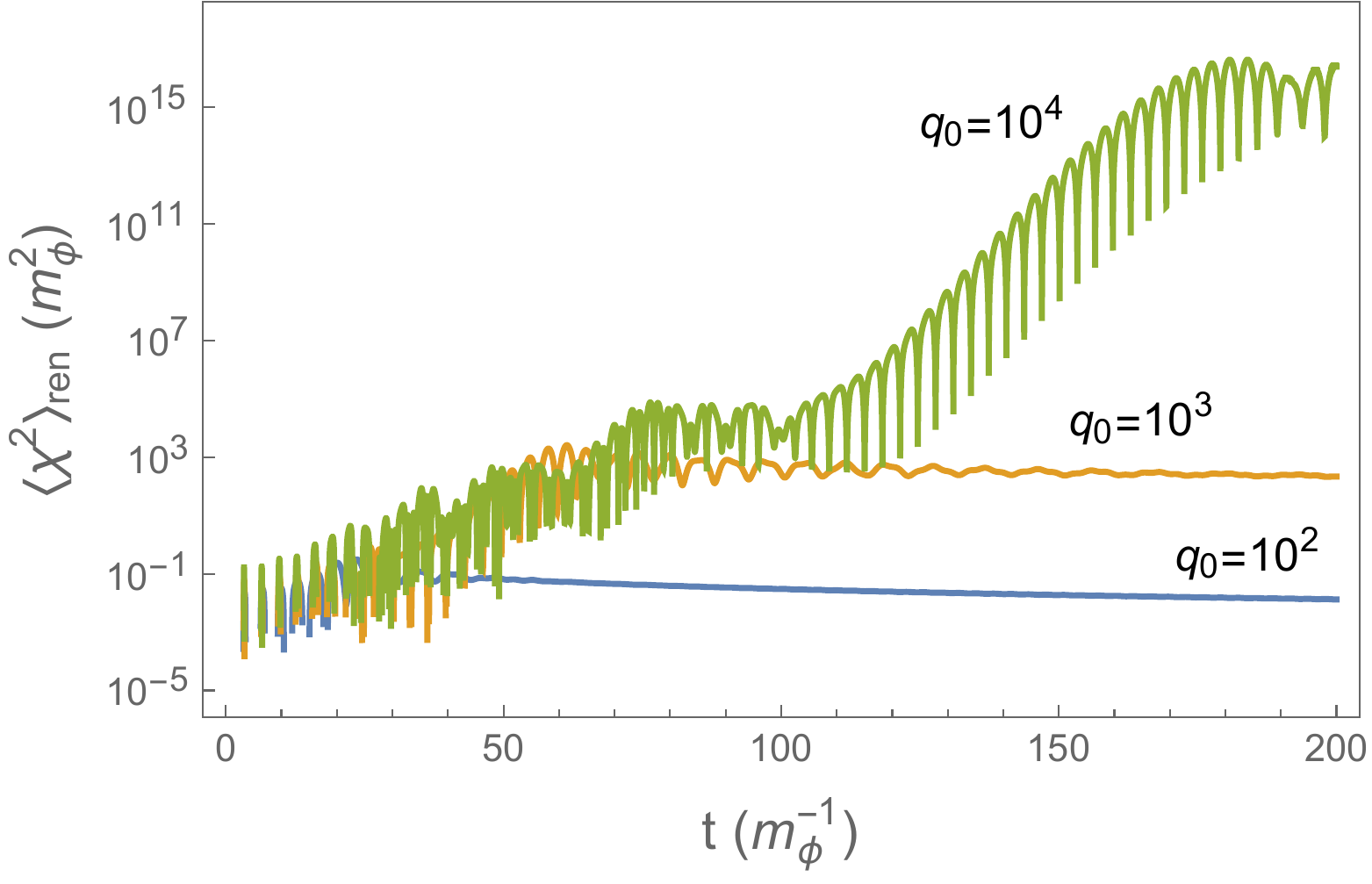}
}
\subfloat[$\lambda_2$ dependence]
{
\includegraphics[width=7.5cm]{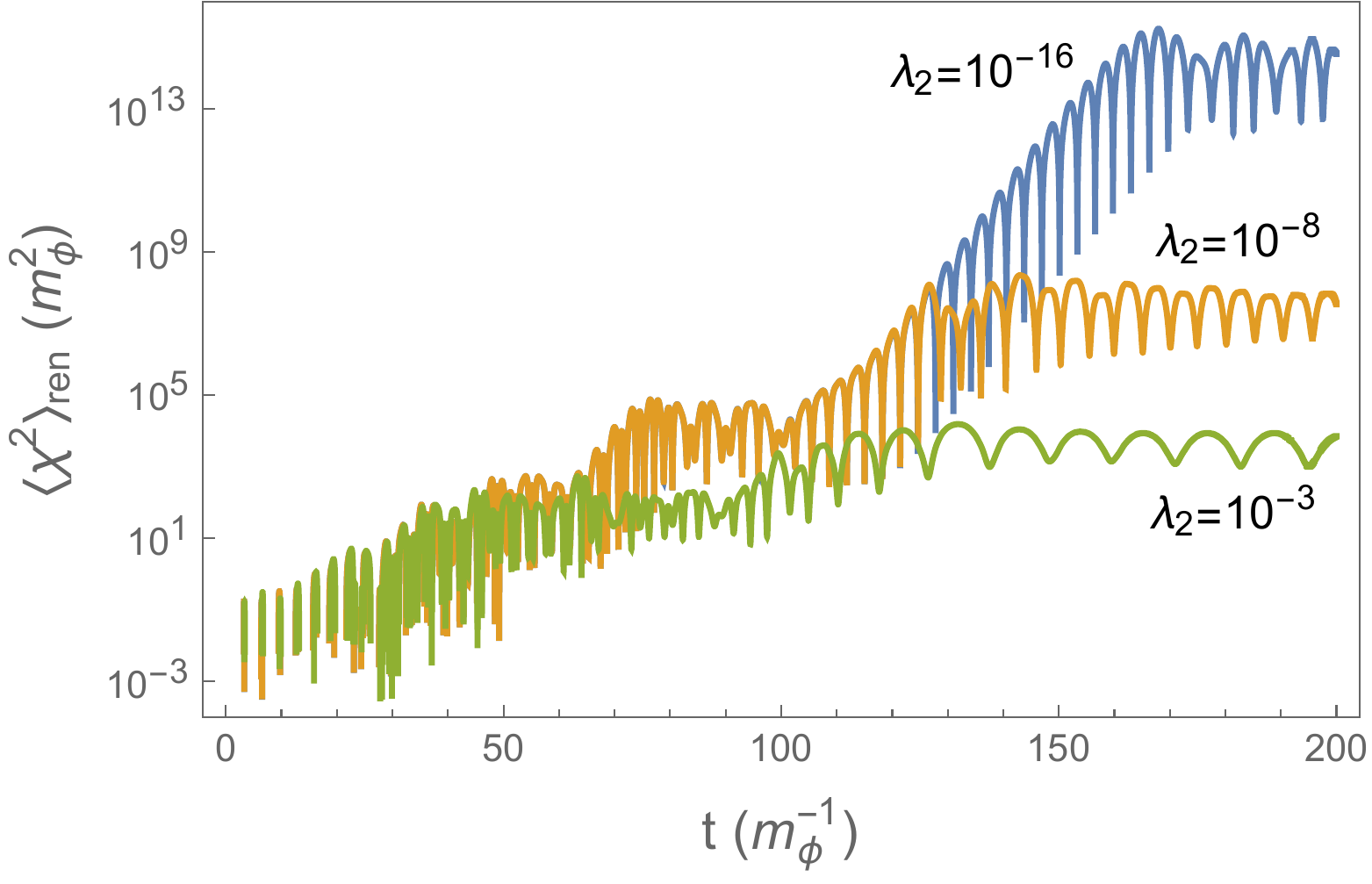}
}
\caption{Panel (a): Evolution of $\braket{\chi^2}_\mathrm{ren}$  for  different values of $q_0$ in the absence of backreaction, $\lambda_2=0$. Note the strong dependence on $q_0$ (the vertical axis is logarithmic.) Panel (b) : Dependence of $\braket{\chi^2}_\mathrm{ren}$ on the coupling constant $\lambda_2$ for  fixed  $q_0=10^4$. Note that, in our units and in the absence of backreaction, the inflaton oscillates with a period of $2\pi$ around the minimum of its potential.   \label{fig:chisqdependences} } 
\end{figure}

As we discuss in appendix  \ref{sec:Equations of Motion}, in the absence of backreaction on the inflaton motion, the spectrum of gravitational waves essentially depends on the single dimensionless parameter
\begin{equation}\label{eq:q0}
	q_0\equiv\frac{2}{3} \frac{\lambda_2  M_P^2}{m_\phi^2},
\end{equation}
which  determines how efficient parametric resonance is. Adding backreaction on the inflaton motion  introduces yet another dimensionless parameter, $\lambda_2$, which controls how $\braket{\chi^2}$ impacts the inflaton motion.   Although $\lambda_2$ and $q_0$ are related by equation (\ref{eq:q0}), it is useful to regard them as independent parameters. The limit $\lambda_2\to 0$ while $q_0$ remains finite, in particular, is the limit in which there is no backreaction of matter on the inflaton motion. Figure \ref{fig:chisqlambda}  displays the dependence of  $\braket{\chi^2}$  on the cutoff used in our numerical implementation in that limit: Panel (a) shows the unrenormalized value of $\braket{\chi^2}$, that is, the cut off integral in (\ref{eq:unren chisq}) computed numerically. Panel (b) shows its renormalized value, that is, the term in square brackets in equation (\ref{eq:ren driving}). Clearly, the renormalized expectation on the right does not depend on $\Lambda$, as expected. In fact, provided that the cutoff $\Lambda$ is not chosen to be much larger than the magnitude of the modes expected to undergo parametric resonance, the subtraction terms needed to renormalize the expectation value play an important role only when $q_0$ is sufficiently small \cite{Armendariz-Picon:2019csc}. Figure \ref{fig:chisqdependences} displays the dependence of $\braket{\chi^2}_\mathrm{ren}$ on the value of $q_0$  in the absence of backreaction, $\lambda_2\to 0$, and  its dependence on $\lambda_2\neq 0$ when  backreaction is taken into account.  In the absence of backreaction  the matter mode functions grow exponentially with $q_0$, and so does  $\braket{\chi^2}_\mathrm{ren}$. This strong dependence, however, is quenched by backreaction, which, as shown on the right panel of figure \ref{fig:chisqdependences},  suppresses $\braket{\chi^2}_\mathrm{ren}$ as the coupling $\lambda_2$ increases. Because of equation (\ref{eq:Veff}), backreaction on the inflaton motion ought to be relevant when $\lambda_2\braket{\chi^2}\gtrsim m_\phi^2$.  This is why  $\braket{\chi^2}_\mathrm{ren}$  stops growing once $\braket{\chi^2}_\mathrm{ren}/m_\phi^2$ becomes of the order of $\lambda_2^{-1}$.

\section{Backreaction on the Metric}
\label{sec:Backreaction on the Metric}

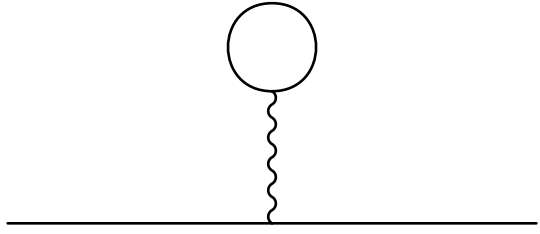
\begin{figure}
\begin{center}
\begin{fmfgraph}(200,100) 
\fmfleft{h1} 
\fmfright{h2}
\fmftop{vt}
\fmf{plain}{h1,vb}
\fmf{plain}{vb,h2}
\fmf{wiggly,tension=0}{vb,vt}
\fmf{plain}{vt,vt}
\end{fmfgraph}
\end{center}
\vspace{-1.5cm}
\caption{Graviton-mediated self-energy corrections to the matter propagator at one loop. Solid lines denote matter fields, and the wavy line the graviton. The graviton source is the energy-momentum tensor, which is quadratic in the matter field $\chi$. \label{fig:Self Energy Graviton} } 
\end{figure}

Conceptionally, the derivation of how quantum corrections affect the evolution of the scale factor is not that different from the one that took us to equation (\ref{eq:quantum eom}). Diagrammatically the  inflaton tadpole diagram in figure \ref{fig:Self Energy} is simply replaced by the graviton tadpole in figure \ref{fig:Self Energy Graviton}.   Just like we demanded above that $\braket{\delta\phi}=0$, we shall perturb the metric around its background value,
$
	g_{\mu\nu}=\bar{g}_{\mu\nu}+\delta g_{\mu\nu},
 $
and demand
$
{\braket{\delta g_{\mu\nu}}=0}.
$
As before, at one loop it suffices to  expand the gravitational action to first order in $\delta g_{\mu\nu}$,  which leads to  the quantum corrected gravitational equations
\begin{equation}\label{eq:Einstein eqs}
	M_P^2\bar{G}^{\mu\nu}=\braket{T^{\mu\nu}},
\end{equation}
where $T_{\mu\nu}$ is the stress tensor of matter in the background spacetime $\bar{g}_{\mu\nu}$, and $\braket{\cdots}$ denotes expectation value.  These are of course the equations of semiclassical gravity, which we have thus ``derived" from a quantum perspective.\footnote{Strictly speaking, in order to  arrive at  equation (\ref{eq:Einstein eqs})   we have to assume that the  propagator for the field fluctuations $\delta g_{\mu\nu}$  exists. But in fact, because of diffeomorphism invariance,  the  propagator is  ill-defined,  unless appropriate gauge fixing terms are introduced. We shall return to this issue in appendix \ref{sec:Boundary Terms}.} 

It shall prove convenient to split the energy-momentum momentum tensor into a component that only depends on the background field $\bar\phi$ and  one that depends on $\chi$,
\begin{subequations}
\begin{equation}\label{eq:split}
	T^{\mu\nu}\equiv T^{\mu\nu}_{(\phi)}+T^{\mu\nu}_{(\chi)}.
\end{equation}
This separation is somewhat artificial, because the fields are actually coupled,  and the effective mass of $\chi$ does depend on the background inflaton. In any case, using that $\braket{\delta\phi}=0$ and ignoring again  the  inflaton loop we find
\begin{align}
\braket{T_{(\phi)}^{\mu\nu}}&\equiv \partial^\mu  \bar\phi \partial^\nu \bar\phi-\bar g^{\mu\nu}
	\left(\frac{1}{2}\partial^\rho \bar\phi \, \partial_\rho \bar\phi +\bar{V}\right),
	\\
\braket{T_{(\chi)}^{\mu\nu}}&=\sum_i\left\langle\partial^\mu \chi_i \partial^\nu \chi_i -\frac{\bar g^{\mu\nu}}{2}
	(\partial^\rho \chi_i \partial_\rho \chi_i+\kappa_i \, \chi_i^2)\right\rangle.
\end{align}
\end{subequations}
Because the expectation of the energy-momentum tensor is spatially constant by homogeneity, it only couples  to  metric perturbations with zero spatial momentum. 

\subsection{Evaluation of the Energy Density}
\label{sec:Evaluation of the Energy Density}
The gravitational equations (\ref{eq:Einstein eqs}) are not all independent, since the Einstein and energy-momentum tensors are covariantly conserved. We shall hence restrict our attention to their time-time component, which we shall deem the  ``Friedman" equation
\begin{equation}\label{eq:Friedman}
	\frac{\mathcal{H}^2}{a^2}=\frac{a^2\braket{T^{00}}}{3M_P^2}, \quad
	a^2\braket{T^{00}}\equiv \bar{\rho}_\phi+\rho_\chi.
\end{equation}
This restriction  also simplifies the role played by  boundary terms, as we discuss in appendix \ref{sec:Boundary Terms}. 
The energy density of matter  $\rho_\chi\equiv a^2\langle T^{00}_{(\chi)}\rangle$ can be expanded again in the number of time derivatives. In this case, to capture all the potentially divergent terms we need to go up to four derivatives, 
\begin{subequations}\label{eq:T vev}
\begin{align}
\rho_\chi^{(0)}&=\frac{1}{8(2\pi^2)}\sum_i \sigma_i \left[\frac{\Lambda^4}{a^4}+\frac{\kappa_i \Lambda^2}{a^2}+\frac{\kappa_i^2 }{4}\left(\frac{1}{2}-\log x_i\right)\right],
\label{eq:T000}
\\
\rho_\chi^{(2)}&=\frac{1}{8(2\pi^2)}\sum_i \sigma_i \left[\frac{\Lambda^2 \mathcal{H}^2}{a^4}-\frac{\kappa_i \mathcal{H}^2}{a^2}\left(\frac{4}{3}-\frac{1}{2}\log x_i\right)
-\frac{\dot \kappa_i \mathcal{H}}{2a^2} \left(\frac{5}{3}-\log x_i\right)
+\frac{\dot{\kappa}^2_i}{24a^2 \kappa_i}
\right],
\label{eq:T002}
\\
\rho_\chi^{(4)}&=\frac{1}{8(2\pi^2)}\sum_i \sigma_i \left[
-\frac{\mathcal{H}^4}{60 a^4}-\frac{4 \mathcal{H}^2 \ddot{a}}{15 a^5}-\frac{19\ddot{a}^2}{60a^6}+
\frac{19\mathcal{H}\dddot{a}}{30 a^5}
+\left(
\frac{\mathcal{H}^2 \ddot{a}}{a^5}+\frac{1}{4}\frac{\ddot{a}^2}{a^6}
-\frac{1}{2}\frac{\mathcal{H}\dddot{a}}{a^5}
\right)\log x_i\right],
\label{eq:T004}
\end{align}
\end{subequations}
where we have dropped terms of order $1/\Lambda$, and in (\ref{eq:T004}) also those that vanish in the limit $\kappa_i\to \infty$. These densities are finite as the cutoff $\Lambda$ approaches infinity, if in addition to  conditions (\ref{eq:sigma cond}), the regulator masses satisfy 
\begin{equation}\label{eq:add sigma cond}
\sum_i \sigma_i M_i^4=0.
\end{equation}
 
Although the expressions  in  (\ref{eq:T vev}) remain finite as the cutoff is removed, they depend on the  regulator masses, and hence diverge when the regulators are decoupled. But these  surviving  contributions are of the same form as those  from the additional counterterms
\begin{equation}\label{eq:counterterms 2}
	S_\mathrm{ct}\supset \int d^4x \sqrt{-g}\left[-\delta \Lambda+\frac{\delta M_P^2}{2}R+\delta c_{(1)} R^2+\delta c_{(2)} R_{\mu\nu} R^{\mu\nu} \right],
\end{equation}
which  renormalize the different terms  in the Friedman equation (\ref{eq:Friedman}). Let us cast the latter in the form
\begin{equation}\label{eq:generalized Friedman}
3M_P^2(\bar\phi) H^2+\cdots=U(\bar\phi)+\frac{1}{2a^2}\dot{\bar\phi}^2+\cdots,
\end{equation}
where the dots stand for terms with higher derivatives of the scalar or the metric. 
At tree level we know that $U(\bar\phi)=\bar V$, but at one loop the quantum corrections in (\ref{eq:T000}) modify this expression. We cancel the divergent $\bar\phi$-independent vacuum  energy piece in $U$ by choosing
\begin{equation}\label{eq:delta Lambda}
	2\pi^2 \delta \Lambda=\frac{1}{32}
	\sum_r\sigma_r M_r^4 \log X_r+\delta\Lambda^f.
\end{equation}
 The dependence of $\delta\Lambda$ on the fourth power of the regulator masses is an expression of the cosmological constant problem.   Because of  relation (\ref{eq:kappai}), equation (\ref{eq:T000}) also yields   regulator-dependent contributions to $U(\bar\phi)$ proportional to $\bar\lambda$ and $\bar\lambda^2$. These   are canceled precisely by the   counterterms proportional to  $\delta d_1$ and $\delta d_2$ in equations (\ref{eq:phi counterterms}). Therefore  the renormalized potential density is 
\begin{equation}\label{eq:Uren}
	\bar U_\mathrm{ren}=\bar V+
	\frac{1}{2\pi^2}\left[
	\frac{(M_0^2+\bar\lambda)^2}{32}\log\frac{M_0^2+\bar\lambda}{M_0^2+\lambda(\mu)}
	+\delta\Lambda^f
	+\left(\delta d^f_1-\frac{M_0^2}{32}\right)\bar\lambda+\left(\delta d^f_2-\frac{3}{64}\right)\bar\lambda^2
	\right].
\end{equation}  
This result illustrates the importance of identifying the counterterms in the theory. If we simply subtract from the  different expectation values their cutoff dependent contributions, there is no guarantee that the undetermined finite pieces in the potential density will be consistent with those of the driving term.   In fact,  in spite of their different origin, at zero derivatives both  are related by
\begin{equation}
	\left\langle \frac{\partial V}{\partial \phi}\right\rangle^{(0)}_\mathrm{ren}
	= \frac{d\bar U_\mathrm{ren}}{d\bar\phi}.
\end{equation}
This property also holds for the regularized but unrenormalized forms of the driving term and the energy density, and trivially applies to the contribution of the counterterms.

The surviving summands  proportional to $\mathcal{H}^2$ in  equation (\ref{eq:T002})  are proportional to the left-hand-side of the Friedman equation (\ref{eq:generalized Friedman}), and thus renormalize the value of the Planck mass. In order to keep the Planck mass finite, the counterterm $\delta M_P^2$ needs to be
\begin{equation}\label{eq:delta M_p}
2\pi^2 \delta M_P^2 =\frac{1}{48}
	\sum_r\sigma_r M_r^2 \log X_r+(\delta M_P^2)^f.
\end{equation}
Because $\kappa_i$ in equation (\ref{eq:T002}) depends on the inflaton, there is also a $\bar\phi$-dependent divergent renormalization of Planck's mass, which is cancelled by the counterterm proportional to $\delta\xi$ in  (\ref{eq:delta xi}).  As a result, the renormalized, inflaton-dependent Planck mass in the effective Friedman equation is
\begin{equation}\label{eq:phi Mp}
	M_P^2(\bar\phi)=M_P^2+\frac{1}{2\pi^2}\left\{(\delta M_P^2)^f-2\delta\xi^f\bar{\lambda}
	+\frac{1}{48}\left[(M_0^2+\bar\lambda)\log\frac{M_0^2+\bar\lambda}{M_0^2+\lambda(\mu)}-\bar{\lambda}\right]\right\},
\end{equation}
which shows that the Planck mass ``runs" with the inflaton value.  Equation (\ref{eq:T002}) also contains terms proportional to $\mathcal{H}\dot{\kappa}_i=\mathcal{H}\dot{\bar\lambda}$. Their divergent contribution cancels the one we obtain by varying the  $\delta\xi$ counterterm in equation (\ref{eq:counterterms 1}), provided that $\delta\xi$ is still given by equation (\ref{eq:delta xi}). There is however a  left-over finite piece  that can be thought of as the renormalized value of the energy density at two derivatives,
\begin{equation}
	\left(\delta\rho_\chi^{(2)}\right)_\mathrm{ren}=
	\frac{1}{2\pi^2}
	\left[6\delta\xi^f-\frac{1}{16}\log \frac{M_0^2+\bar{\lambda}}{M_0^2+\lambda(\mu)}\right]\frac{\mathcal{H}\dot{\bar\lambda}}{a^2}.
\end{equation}

In order to determine the values of the counterterms proportional to $\delta c_{(1)}$ and $\delta c_{(2)}$ in equation (\ref{eq:counterterms 2})   we need to find out how they contribute to the gravitational equations. Their variation with respect to the metric yields the  tensors usually labeled ${}^{(1)}\!H_{\mu\nu}$ and  ${}^{(2)}\!H_{\mu\nu}$ respectively. A relatively direct route  to easily compute their $00$ components   is to substitute the ADM metric ${ds^2=-N^2 dt^2 +a^2(t) d\vec{x}\,^2}$ into the action (\ref{eq:counterterms 2}), vary with respect to the lapse function $N$, and then choose a gauge with $N=a$. The result is 
\begin{subequations}
\begin{align}
\frac{1}{\mathcal V}\frac{\delta }{\delta N} \left(\int d^4x \sqrt{-g}\, R^2 \right)&=
-144\frac{\dot{a}^2\ddot{a}}{a^4}-36\frac{\ddot{a}^2}{a^3}+72 \frac{\dot{a}\,\dddot{a}}{a^3},
\\
\frac{1}{\mathcal V}\frac{\delta }{\delta N} \left(\int d^4x \sqrt{-g}\, R_{\mu\nu}R^{\mu\nu} \right)&=-48\frac{\dot{a}^2\ddot{a}}{a^4}-12\frac{\ddot{a}^2}{a^3}+24 \frac{\dot{a}\,\dddot{a}}{a^3},
\end{align}
\end{subequations}
where we have ignored boundary terms, which we discuss in appendix \ref{sec:Boundary Terms}. The two variations are proportional to each other, owing to the fact that in a conformally flat spacetime $ {}^{(1)}\!H_{ab}=3 \times\!{}^{(2)}\!H_{ab}$. The log  divergent piece in $\langle \rho_\chi\rangle^{(4)}$ is also proportional to either variation. Therefore,  by demanding  the cancellation of this divergence, we can only fix a linear combination of $\delta c_{(1)}$ and $\delta c_{(2)}$,
\begin{equation}\label{eq:delta c1 and c2}
	2\pi^2 (3\delta c_{(1)}+\delta c_{(2)})=-\frac{1}{384}
	\sum_r\sigma_r \log X_r+\delta c^f.
\end{equation}
To disentangle the two individual contributions we would have to consider  more general backgrounds.   In any case,  the renormalized  energy density at four derivatives thus becomes
\begin{equation}
	\delta\rho_\chi^{(4)}=\frac{1}{2\pi^2}
	\left[48\,\delta c^f-\frac{1}{8}\log \frac{M_0^2+\bar\lambda}{M_0^2+\lambda(\mu)}\right]\left(\frac{\mathcal{H}^2\ddot{a}}{a^5}
	+\frac{1}{4}\frac{\ddot{a}^2}{a^6}-\frac{1}{2}\frac{\mathcal{H}\dddot{a}}{a^5}\right).
\end{equation}

 Finally, with the counterterms given by the previous expressions, the  one-loop ``exact" renormalized value of the energy density reads
 \begin{subequations}\label{eq:T00 final}
\begin{equation}
\begin{split}
	\braket{\rho_\chi}_\mathrm{ren}&= \frac{1}{2\pi^2}  \Bigg\{\frac{1}{2a^2}
	\lim_{\Lambda\to\infty}
	\int_0^\Lambda dk \, k^2 \left[\dot{w}_k \dot{w}^*_k+(k^2+\kappa_0 a^2)w_k w^*_k\right]
	\\
	&-\frac{\Lambda^4}{8a^4}
	-\frac{\kappa_0 \Lambda^2}{8a^2}
	+\frac{\kappa_0^2}{32}\log \frac{4\Lambda^2}{a^2 \kappa_\mu}	
	+f_0
	\\
	 &-\frac{\Lambda^2\mathcal{H}^2}{8a^4}-\frac{\kappa_0 \mathcal{H}^2}{16a^2}\log \frac{4\Lambda^2}{a^2 \kappa_\mu}	
	 -\frac{\dot{\bar\lambda} \mathcal{H}}{16a^2}
	 	\log \frac{4\Lambda^2}{a^2 \kappa_\mu}
	+f_2
	  \\
	 &-\frac{1}{8}\left(\frac{\mathcal{H}^2 \ddot{a}}{a^5}+\frac{1}{4}\frac{\ddot{a}^2}{a^6}-\frac{1}{2}\frac{\mathcal{H}\dddot{a}}{a^5}
\right)\log\frac{4\Lambda^2}{a^2\kappa_\mu } +f_4
	 \Bigg\},
\end{split}
\end{equation}
where  the remaining cutoff independent finite contributions are
\begin{eqnarray}
	f_0&\equiv& \delta\Lambda^f-\frac{M_0^4}{64}
	+\left(\delta d_1^f-\frac{M_0^2}{16}\right)\bar{\lambda}
	+\left(\delta d_2^f-\frac{1}{16}\right)\bar{\lambda}^2, 
	\\
	f_2&\equiv&\frac{\mathcal{H}^2}{a^2}\left[\frac{M_0^2}{6}-3(\delta M_P^2)^f
		+\left(\frac{11}{48}+6\delta\xi^f\right)\bar\lambda\right]
		+\frac{\mathcal{H}\dot{\bar\lambda}}{a^2}\left[6\delta\xi^f+\frac{5}{48}\right],
	\\
	f_4&\equiv & \frac{\mathcal{H}^4}{480 a^4}
	+\frac{\mathcal{H}^2 \ddot{a}}{30 a^5}+\frac{19\ddot{a}^2}{480a^6}
	-\frac{19\mathcal{H}\dddot{a}}{240 a^5}
	+48 \delta c^f \left(\frac{\mathcal{H}^2 \ddot{a}}{a^5}+\frac{1}{4}\frac{\ddot{a}^2}{a^6}-\frac{1}{2}\frac{\mathcal{H}\dddot{a}}{a^5}
\right).
\end{eqnarray}
\end{subequations}
We compute the  renormalized matter pressure along the same lines in appendix \ref{sec:Pressure Renormalization}.

It can be  checked explicitly that  (\ref{eq:T00 final}) yields  the renormalized expressions that we have derived above. Although the regulators play no role in this final expression, the renormalized energy density does not simply follow from removing the cutoff dependence from the mode integral. Our regularization scheme forces us to subtract a term proportional to $\mathcal{H}^4/a^4$, but the latter does not appear in any of the log-divergent terms. In addition,  the finite pieces at any number of derivatives do not have the same structure as the corresponding log-divergent terms. This again underscores the importance of a complete analysis of regularization and renormalization. Removing the cutoff-dependent contributions by hand, or simply isolating the logarithmic divergences is not sufficient. 

The reader may have  recognized in equation (\ref{eq:T00 final}) an expression akin to the renormalized energy-momentum tensor in the adiabatic scheme. This is because in the latter one subtracts from the divergent expectation an expansion of the same quantity in the number of time derivatives, up to the number that renders the expression finite (four in the case of $T^{\mu\nu}$). Apart from  terms that vanish when $\kappa_i\to\infty$, and we did not write down, this expansion is precisely the one contained in equations (\ref{eq:T vev}) when restricted just to the $i=0$ contribution. This   also  happens to be essentially  what we subtract in the renormalized expression (\ref{eq:T00 final}). Thus, the main differences between the adiabatic scheme and Pauli-Villars   are that  $i)$   expression (\ref{eq:T00 final}) also contains the explicit contributions from the counterterms needed to enforce the renormalization conditions, $ii)$ in  the adiabatic scheme one would presumably subtract $\phi$-dependent logarithms, $\log  \kappa_0$ instead of  the $\log \kappa_\mu$ in (\ref{eq:T00 final}), $iii)$ we are not subtracting any of the time derivatives of $\bar{\phi}$ that arise from the dependence of the effective square mass  $\kappa_0$ on the inflaton $\bar\phi$. These differences are inconsequential for free fields, but are significant when their mass depends on an external field like $\bar\phi$.  In the former case, our derivation can be thought of as further justification for the standard adiabatic scheme. In general however, we feel that Pauli-Villars is a superior regularization scheme. To our knowledge,   its use  in cosmology  was first advocated for by S.~Weinberg in reference \cite{Weinberg:2010wq}.

\subsection{Order Reduction}

The higher-curvature corrections needed to  renormalize the theory  in equation (\ref{eq:T00 final}) should be appropriately interpreted. In this work we regard general relativity as a low-energy effective field  theory, and, as such, the renormalized contributions of the counterterms   are only supposed to capture corrections to the equations of motion derived at lowest order, that is, in general relativity. 

The  higher time derivatives present in the corrected Friedman equation  (\ref{eq:Friedman}), however, suggest otherwise. They not only imply that the order of this differential equation changes, which amounts to a change in the number of degrees of freedom in the theory, but also  lead to solutions that cannot be thought of as corrections to the lowest order theory. An illuminating example is that of Starobinsky inflation \cite{Starobinsky:1980te}, in which a literal solution of the equations of motion in the presence of an  $R^2$ term yields inflating solutions that are  altogether absent  in general relativity \cite{Simon:1991bm}.

An appropriate way to deal with higher derivative corrections from the point of view of  effective field theory has been lucidly discussed by J.~Z.~Simon in  \cite{Simon:1990ic}. In practice, it  implies that we can use the ``tree-level"  Friedman equation  to reduce the order of the quantum-corrected equation (\ref{eq:Friedman}), as explicitly discussed in \cite{Parker:1993dk}. This procedure almost guarantees that the solutions of the reduced equations display the desired property of being low-energy expansions grounded on the lowest order approximation of general relativity. 

\begin{figure}
\subfloat[Unrenormalized]
{
\includegraphics[width=7.5cm]{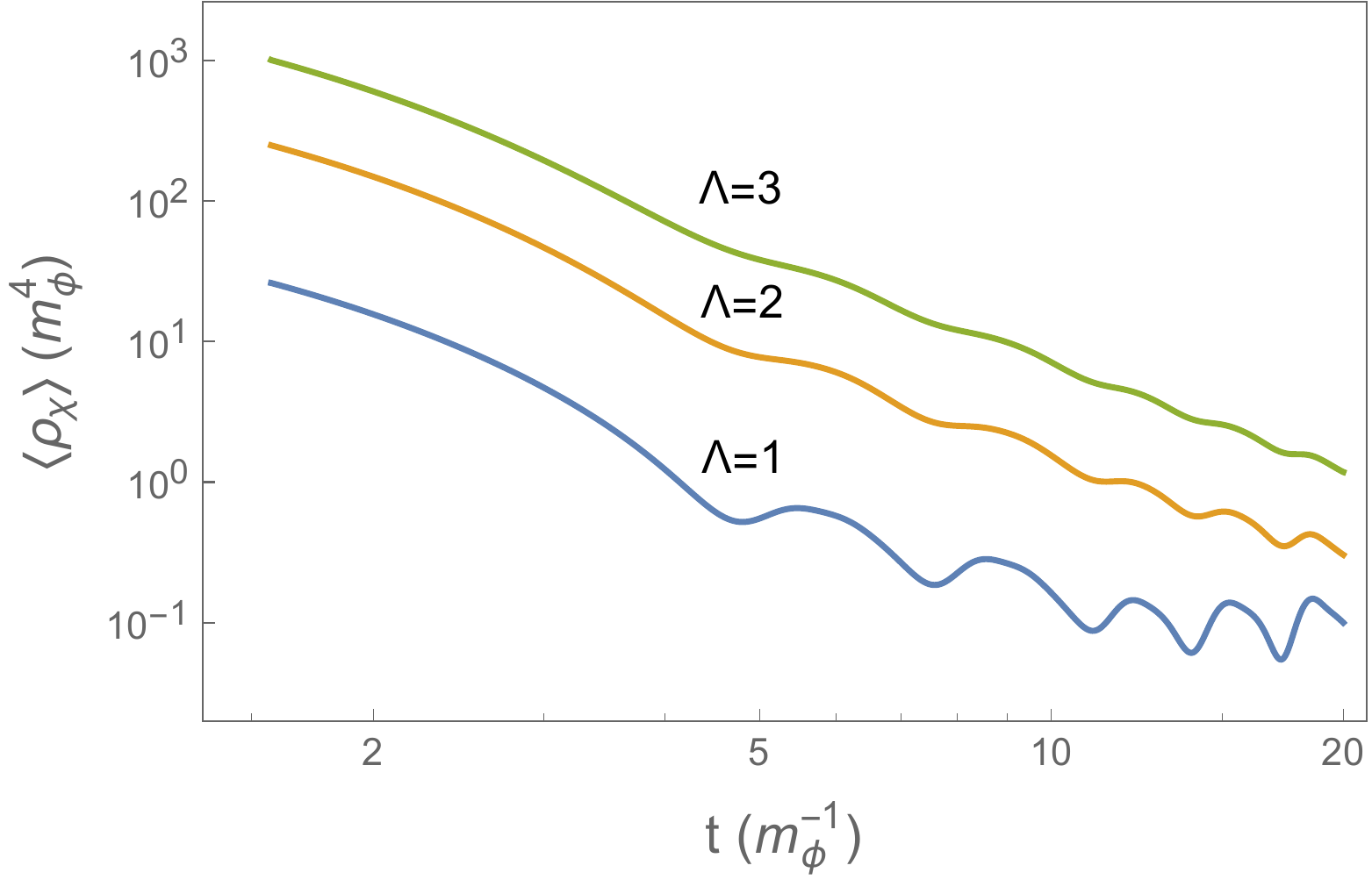}
}
\subfloat[Renormalized]
{
\includegraphics[width=7.5cm]{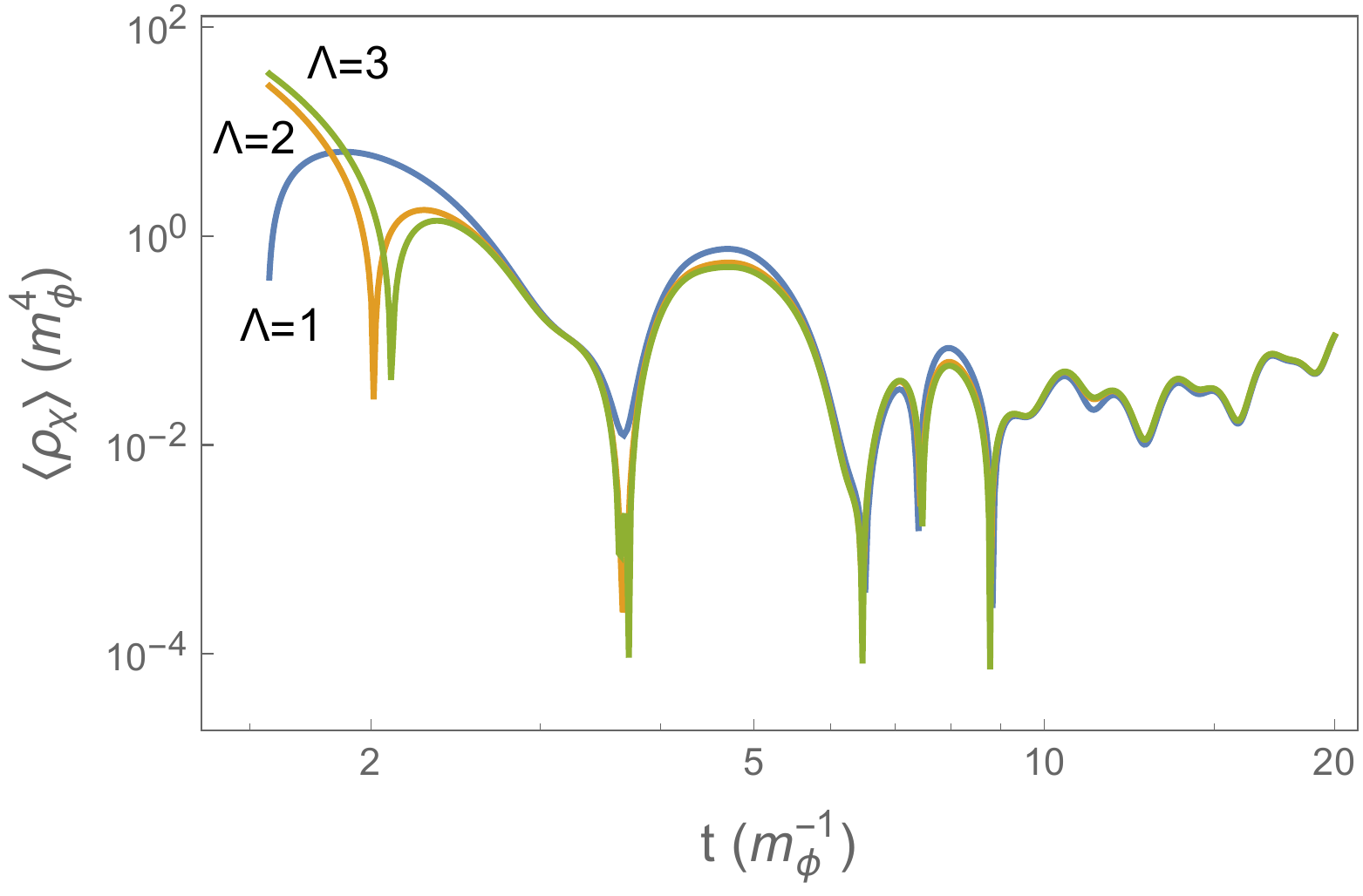}
}
\caption{Evolution of $\braket{\rho_\chi}$  for $q_0=10^2$ in the absence of inflaton and metric backreaction ($\lambda_2=r=0$).  Panel (a) shows again a dependence of the  unrenormalized expectation on the cutoff $\Lambda$. In Panel (b), the dependence of the renormalized expectation  is seen to essentially cancel out, as the three curves become almost indistinguishable.   As in the case of $\braket{\chi^2}$, the unrenormalized energy density is positive by construction, but its renormalized counterpart can take negative values. \label{fig:rhoLambda} } 
\end{figure}

Reference  \cite{Parker:1993dk} carried out such an  order reduction in a universe dominated by radiation, but not in the case of a scalar-field dominated universe that occupies us. To reduce  the order of the ``Friedman" equation (\ref{eq:Friedman}) in that case we shall simply use the $i$-$j$ Einstein equation 
\begin{equation}\label{eq:Einsteinij} 
\frac{\ddot a}{a^3}=\frac{1}{6 M_P^2}\left(-\dot{\bar\phi}^2+4 \bar V a^2 \right)
\end{equation}
and its time derivative to replace $\ddot{a}$ and $\dddot{a}$ in equation (\ref{eq:T00 final}) by expressions with a lower number of derivatives. This procedure leaves factors of  $\ddot{\bar\phi}$, which we eliminate using the inflaton equation of motion. The  outcome is that  the last line in equation  (\ref{eq:T00 final}) should be replaced by 
\begin{equation}
\begin{split}
	\frac{\mathcal{H}^4}{480 a^4}
	&-\frac{\mathcal{H}^2 (196 a^2V+65\dot\phi^2)}{1440 M_P^2 a^4}
	-\frac{19 \mathcal{H}\bar{V}'\bar{\phi}'}{240 M_P^2 a^2}
	+\frac{19 (\dot{\bar\phi}^2)^2-4a^2 \bar V}{17280 M_P^4 a^4}
	\\&
	+\left(48\delta c^f-\frac{1}{8}\log\frac{4\Lambda^2}{a^2\kappa_\mu }\right)\left[
	\frac{(\dot{\bar\phi}^2-4a^2 \bar V)^2}{144 M_P^4 a^4}
	-\frac{\mathcal{H} \bar{V}' \dot{\bar\phi}}{2M_P^2 a^2}
	-\frac{\mathcal{H}^2(5\dot\phi^2+4a^2 V)}{12 M_P^2 a^4}
	\right]
	,
\end{split}
\end{equation}
which manifestly preserves the differential order of the original Friedman equation. 

These considerations also affect the evaluation of the renormalized energy density (\ref{eq:T00 final}). In that  equation, $\mathcal{H}$  refers to the  expansion rate at the corresponding  time $t$,  which is putatively constrained   by equation (\ref{eq:Friedman}). But since $\braket{\rho_\chi}_\mathrm{ren}$ captures the lowest-order correction in the effective field theory expansion,  $\mathcal{H}$ in equation (\ref{eq:T00 final}) ought to be evaluated using the zeroth order Friedman equation $3M_P^2 \mathcal{H}^2=a^2 \bar\rho_\phi$.

\subsection{Numerical Results}
\label{sec:Backreaction Numerical}
Our numerical solutions also include backreaction on the expansion of the universe.   For illustration, we shall return again to the specific quadratic potential and coupling function (\ref{eq:example}), with the finite values of the counterterms given by equation (\ref{eq:finite CT 1}).  To fix the value of the  remaining counterterms we set $\delta c^f=0$ and  demand  that $M_P^2(\mu)=M_P^2$ and  $U_\mathrm{ren}(\bar\phi=0)=0$, which imply that 
\begin{equation}\label{eq:finite CT 2}
	\delta\Lambda^f=0, \quad (\delta M_P^2)^f=0.
\end{equation}

As detailed in appendix \ref{sec:Equations of Motion}, in this case the  impact of backreaction on the metric is controlled by the dimensionless ratio 
\begin{equation}\label{eq:r}
	r\equiv \frac{m_\phi^2}{M_P^2}=\frac{2\lambda_2}{3q_0}
\end{equation}
which, in spite of its relation to $\lambda_2$ and $q_0$, we regard as an independent parameter in our numerical code. In the limit $r\to 0$ metric backreaction is turned off, whereas as $r$ grows it becomes increasingly important.

 \begin{figure}
\subfloat
{
\includegraphics[width=7.5cm]{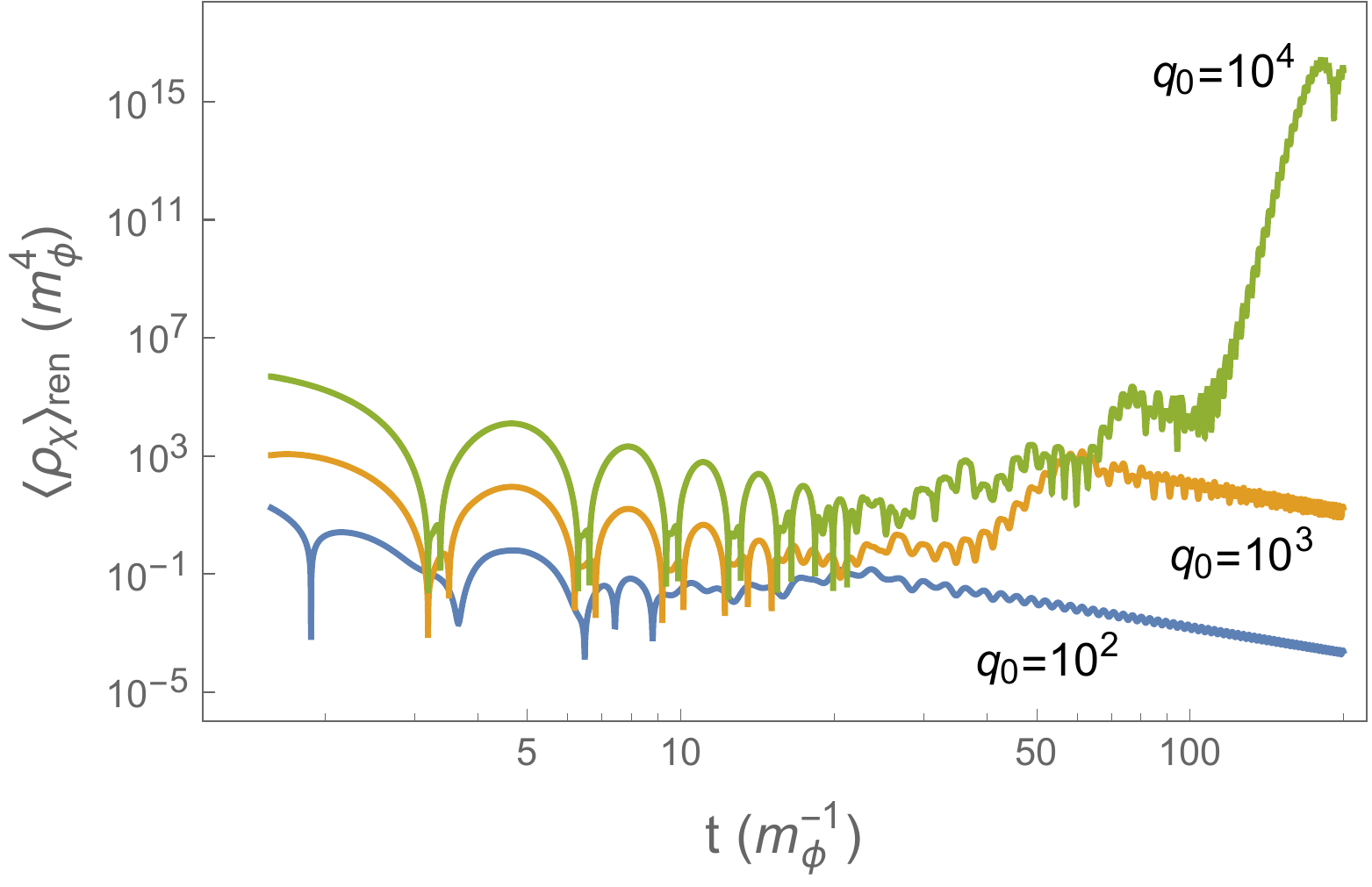}
}
\subfloat
{
\includegraphics[width=7.5cm]{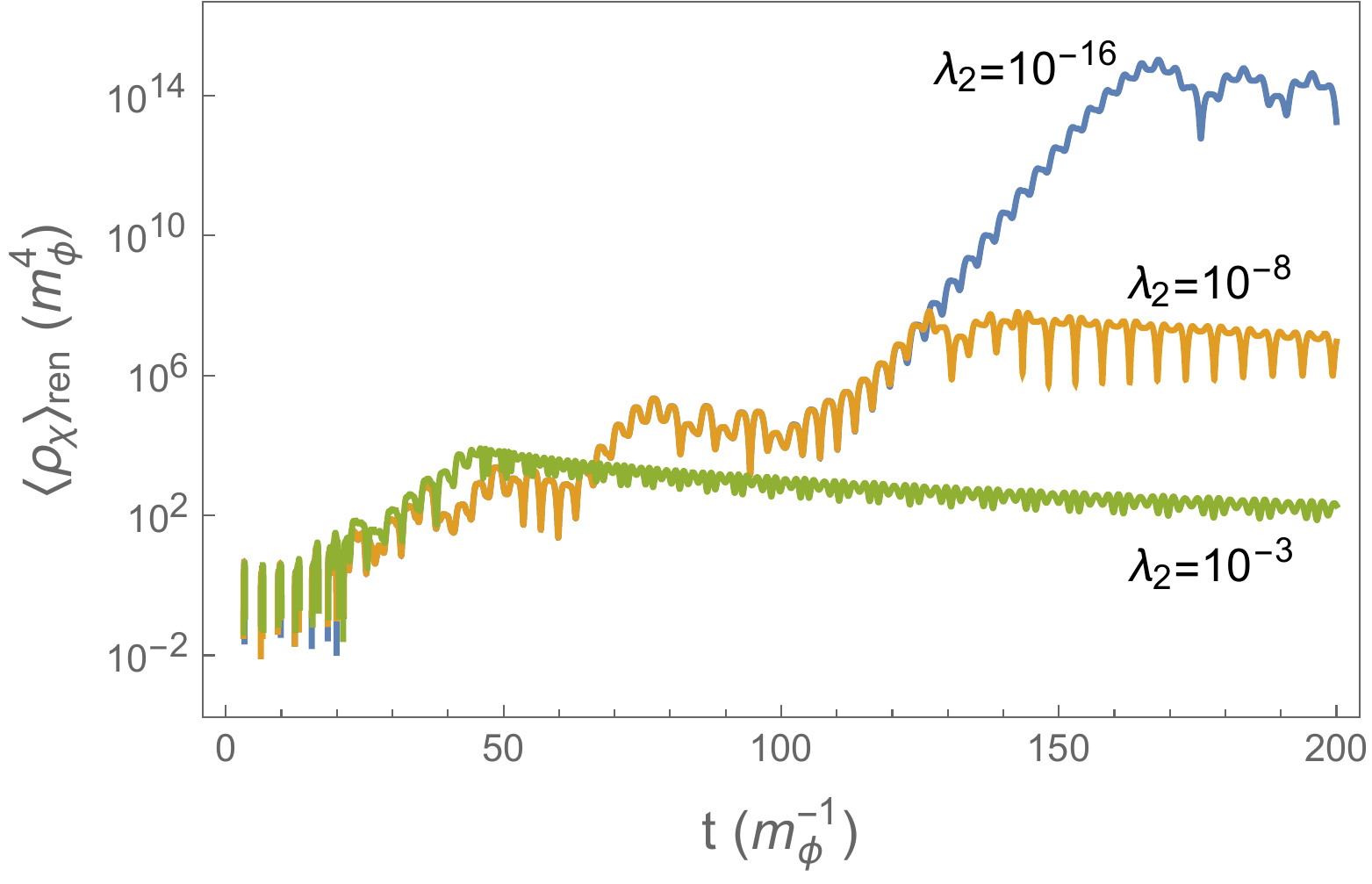}
}
\caption{Left panel: Evolution of $\braket{\rho_\chi}_\mathrm{ren}$  for  different values of $q_0$ in the absence of backreaction, $\lambda_2=0, r=0$. After an  initial period of growth, the density redshifts as $x^{-8/3}$, as expected from radiation. Right panel: Dependence of $\braket{\rho_\chi}_\mathrm{ren}$ on the coupling constant $\lambda_2$ for  fixed  $q_0=10^4$ and $r$ given by equation (\ref{eq:r}).   As $\lambda_2$ increases, backreaction becomes more important, thus limiting the growth of the mater density.
\label{fig:rhodependences} } 
\end{figure}

Figure \ref{fig:rhoLambda} shows the dependence of both the unrenormalized and renormalized matter energy density for different values of the cutoff  in the absence of backreaction, both on the inflaton and on the expansion. The independence of the renormalized expectation on the value of the cutoff is a powerful check not only of our renormalization procedure, but also on the numerical algorithm itself. Note in particular that the renormalization of $\braket{\rho_\chi}$ requires the subtraction of a term of order $\Lambda^4$, whereas that of $\braket{\chi^2}$ only requires the subtraction of a term of order $\Lambda^2$.  The dependence of $\braket{\rho_\chi}_\mathrm{ren}$ on $q_0$ and $\lambda_2$ is explored in figure \ref{fig:rhodependences}. Parametric resonance is only effective during the first oscillations of the inflaton, and ceases to be effective after a time proportional to  $\sqrt{q_0}$ \cite{Kofman:1997yn}. Panel (a) of figure  \ref{fig:rhodependences}  also shows how  the energy density  of matter redshifts  after parametric resonance becomes ineffective.   The $x^{-8/3}$ decay  indicates that the latter behaves as radiation, as expected from our choice $M_0=0$.\footnote{As the inflaton oscillates about its minimum, the scale factor grows as in a matter-dominated universe, $a\propto x^{2/3}$, where $x=m_\phi\tau$ is cosmic time in units of the inverse inflaton mass. Therefore radiation is expected to redshift like $a^{-4}\propto x^{-8/3}$.}

\begin{figure}
\subfloat[Inflaton density] 
{
\includegraphics[width=7.5cm]{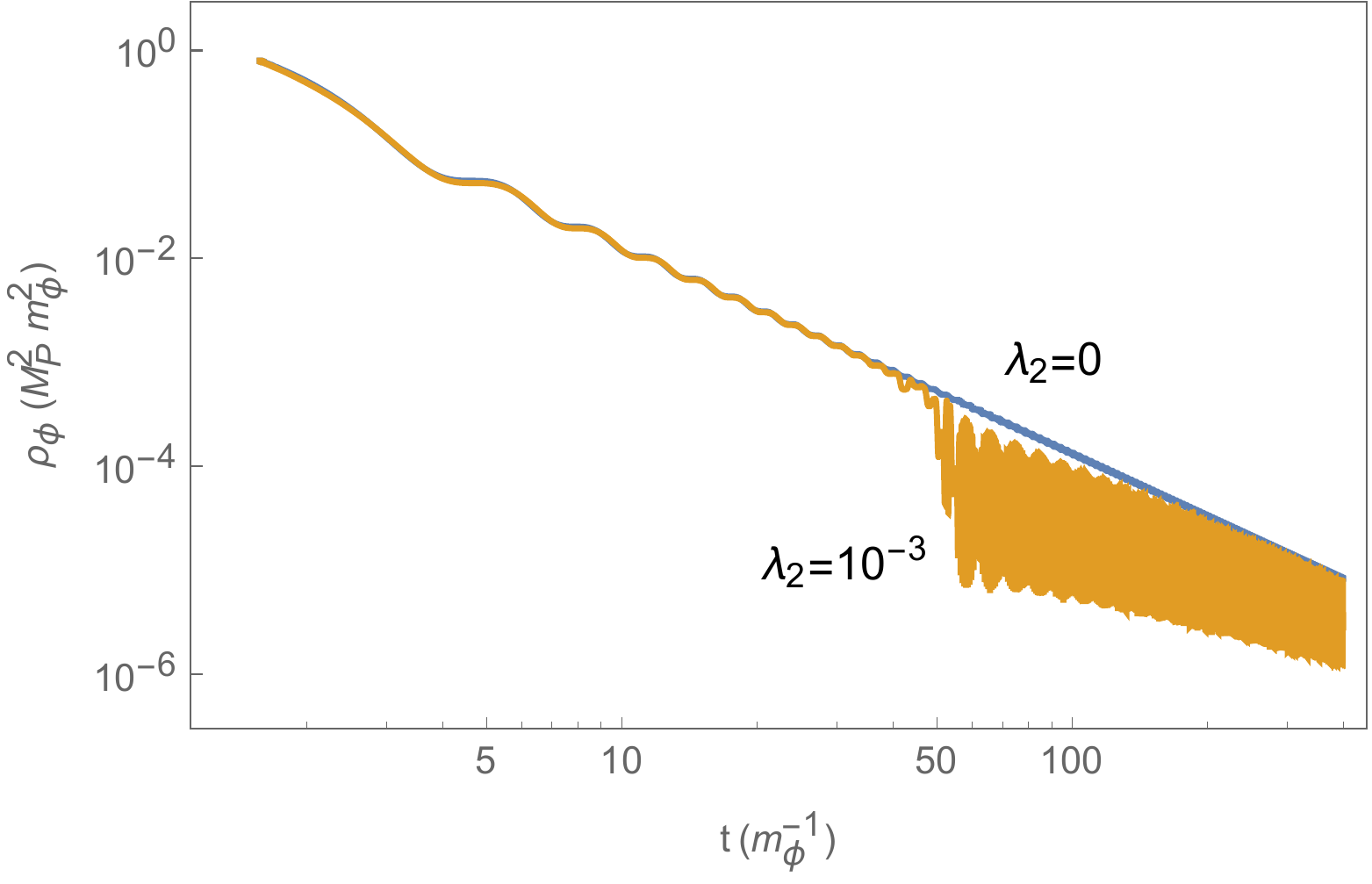}
}
\subfloat[Scale factor]
{
\includegraphics[width=7.5cm]{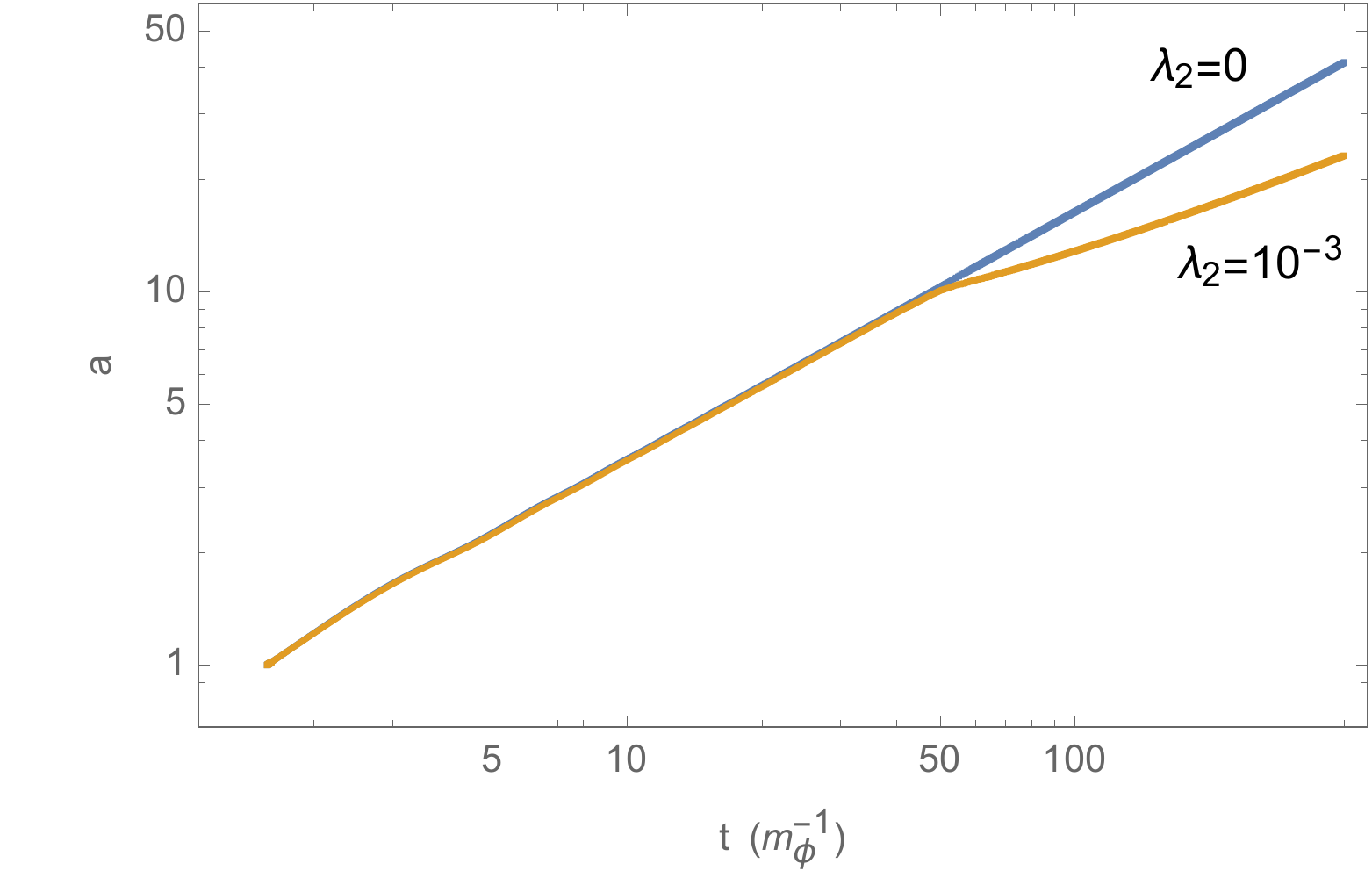}
}
\caption{Evolution of $\bar\rho_\phi$  and the scale factor for   $q_0=10^4$ and $\lambda=0,10^{-3}$ with  backreaction on the metric excluded, $r=0$. Panel (a): The coupling to matter clearly causes a drop in the energy density, which keeps oscillating with a larger amplitude and a suppressed mean value.  Panel (b):  Since $r=0$, the scale factor evolution is only determined by $\bar{\rho}_\phi$. When $\lambda_2=0$  the scale factor grows as in a matter-dominated universe ($a\propto x^{2/3}$). Interestingly, however, when $\lambda_2=10^{-3}$ the scale factor changes behavior after $t\approx 50 \, m_\phi$ and its growth is fitted instead by $a\propto x^{0.4}$. This implies that the backreaction on the inflaton motion causes its density to decay faster than radiation. 
\label{fig:RhoPhi} } 
\end{figure}

The effects of backreaction on the inflaton motion are illustrated in figure \ref{fig:RhoPhi}. The evolution of the inflaton is seen to depart from the one in the absence of backreaction precisely around   the time the density of matter reaches its maximum (see panel (b) in figure \ref{fig:rhodependences}.) We interpret this departure as the decay of the inflaton into radiation. The inflaton density subsequently shows a strong oscillatory behavior  with an average value that steadily decreases. To determine the effective equation of state of  the inflaton during such period, we plot the scale factor as a function of time in panel (b) with the parameter $r$ set to zero (matter fields then have no impact on the universe expansion.) There is a sharp break in the evolution of $a$, from matter-dominated expansion to that associated with an effective equation of state ``stiffer" than that of radiation, $w_\phi\approx 2/3$. The latter suggests that  radiation  can come to dominate the universe. 

Because cosmic expansion is dictated by the total energy density, a simple way to determine whether backreaction on cosmic expansion is relevant is to compare the energy density of matter with that of the inflaton. In fact,  $\braket{\rho_\chi}_\mathrm{ren}$  needs to surpass  $\bar{\rho}_\phi$ for reheating to be successful. In terms of the relevant dimensionless ratio, this condition becomes
\begin{equation}
	\frac{m_\phi^2}{M_P^2} \frac{\braket{\rho_\chi}_\mathrm{ren}}{m_\phi^4}\gtrsim \frac{\bar\rho_\phi}{m_\phi^2 M_P^2}.
\end{equation}
Since it is reasonable to assume that $m_\phi<M_P$, a necessary condition for backreaction to play a role is that  $\braket{\rho_\chi}_\mathrm{ren}/m_\phi^4$ grow larger than  $\bar\rho_\phi/(m_\phi^2 M_P^2)$. This stronger condition is  useful because the two  dimensionless ratios can be estimated with the value of $q_0$ alone. Our numerical analysis indeed indicates that $\braket{\rho_\chi}_\mathrm{ren}/m_\phi^4$ can become larger than $\bar\rho_\phi/(m_\phi^2 M_P^2)$ only  if $q_0\gtrsim 25$. Thus, within our approximations, it appears that a successful reheating is not possible for smaller values of $q_0$. This conclusion  also  agrees  with the  results of the perturbative analysis of, say,  reference \cite{Kofman:1997yn}.   Equation (\ref{eq:r}) may suggest that an increase in $q_0$ makes backreaction less important, but, in fact, because the matter density grows exponentially with $q_0$, the opposite is true.   Yet successful reheating also demands  that the matter density remain above that of the  inflaton.   After its initial growth  during the initial stages of preheating, the matter density begins to redshift like radiation, whereas in the absence of backreaction that of the inflaton redshifts like non-relativistic matter.   It is  then important  that  backreaction on the inflaton motion, which is controlled by the parameter $\lambda_2$, be large enough to affect its evolution. Since $\lambda_2$ appears in combination with $\braket{\rho_\chi}_\mathrm{ren}$, and the later grows exponentially with $q_0$, it typically suffices that $\lambda_2$ not be too small. As we mentioned earlier, when backreaction on the inflaton motion is relevant, its energy density decays faster than radiation, as required by a successful reheating.

\subsection{Comparison with Previous Approaches}

Our analysis may also cast some light into the validity of  the standard computations, which rely on the solution of the classical field equations in the presence of matter waves. In  previous work \cite{Armendariz-Picon:2019csc} we argued that, in the absence of backreaction, the power spectrum of gravitational waves estimated by averaging over an ensemble of classical simulations  approximately agrees with the predictions of the $in$-$in$ formalism provided that  $i)$ parametric resonance is effective, $ii)$  the mode sums do not extend far beyond those modes that experience parametric amplification, $iii)$ initial matter amplitudes are chosen  in the simulations  to match the statistical properties in the quantum theory. Because the source of gravitational waves is the energy-momentum tensor, conditions $i)$ and $ii)$ imply that the contribution of the counterterms remains negligible, and  condition $iii)$ then warrants that the ensemble average in the simulations \emph{qualitatively} matches the quantum expectation of the free matter fields of the $in$-$in$ formalism.  

Turning to  backreaction, as captured by the quantum-corrected equations of motion (\ref{eq:quantum eom}) and (\ref{eq:Friedman}), we note that the correction terms themselves involve the expectation of expressions quadratic in the matter fields, $\chi^2$ in the case of  the inflaton, and $T^{\mu\nu}_\chi$ in the case of matter.  The standard computations use the volume average of the energy density and pressure as sources in the Einstein equations. Because the former are again quadratic in the fields, we thus expect our quantum-corrected Einstein equations to qualitatively agree with those employed in the simulations, by ergodicity. On the other hand, the inflaton equation of motion that is solved in  the standard simulations contains a coupling of the inflaton to $\chi^2$, not its spatial average. As a result, in these simulations the inflaton develops inhomogeneities, which also contribute to the evolution of the matter field $\chi$ and the effective energy density  and pressure in the Einstein equations.   This is to be contrasted with our quantum-corrected equation of motion, in which the expectation $\braket{\chi^2}$ is spatially constant by symmetry, and it thus suffices to consider an homogeneous inflaton background $\bar{\phi}$. Given the non-linear nature of the evolution, it is hard to assess how this difference affects the ensemble average of the gravitational power spectrum in the simulations. We just note that, as we point out in appendix \ref{sec:Covariant Conservation}, the  Einstein equations are not independent, but are related instead to each other by the field equation of motion. Hence, taking the spatial average in the Einstein equations but not in the field equations may lead to inconsistencies.  

\section{Spectrum of Gravitational Waves}

We finally are in a position to determine how backreaction affects the spectrum of  gravitational waves generated during preheating. To do so we have extended the numerical code developed in \cite{Armendariz-Picon:2019csc} by replacing the equation of motion of the inflaton by  (\ref{eq:quantum eom}), and by substituting the original Friedman equation by (\ref{eq:Friedman}). In addition, because the mode equation for the gravitational waves is  usually derived under the assumption that the background satisfies the classical equations, the former needs to be modified too, although the change has a negligible impact for the parameters we sample.  See appendix \ref{sec:Equations of Motion} for the complete set of additional equations  solved by our numerical implementation of backreaction. 

\subsection{The Starobinsky Model}
In order to illustrate our results, we shall adopt a specific inflationary model. Unfortunately, the ``canonical" quadratic inflationary model that served  as basis for earlier gravitational wave calculations and our own examples has been essentially ruled out by a  combination of BICEP2/Keck Array and Planck collaboration data  \cite{Akrami:2018odb}. Following reference \cite{Armendariz-Picon:2019csc}, and trading   simplicity for phenomenological success, we shall focus instead on the   Starobinsky model. As we mentioned earlier, the $R^2$ formulation of the Starobinsky model is hard to reconcile with an effective field theory interpretation of the theory. We shall hence  study its scalar field incarnation, with potential
\begin{equation}\label{eq:Starobinsky}
	V(\phi)=\frac{3 M_P^2 m_\phi^2}{4} \left[1-\exp\left(-\sqrt{\frac{2}{3}} \frac{\phi}{M_P}\right)\right]^2.
\end{equation}
An advantage of this potential is that we do not need to introduce any additional dimensionless parameters in our analysis. The coupling function $\lambda(\phi)$, the mass of $\chi$, the renormalization scale $\mu$ and  the finite pieces of the counterterms are those in equations (\ref{eq:example}), (\ref{eq:masses}), (\ref{eq:finite CT 1}) and (\ref{eq:finite CT 2}).

\begin{figure}
\begin{center}
\includegraphics[height=6cm]{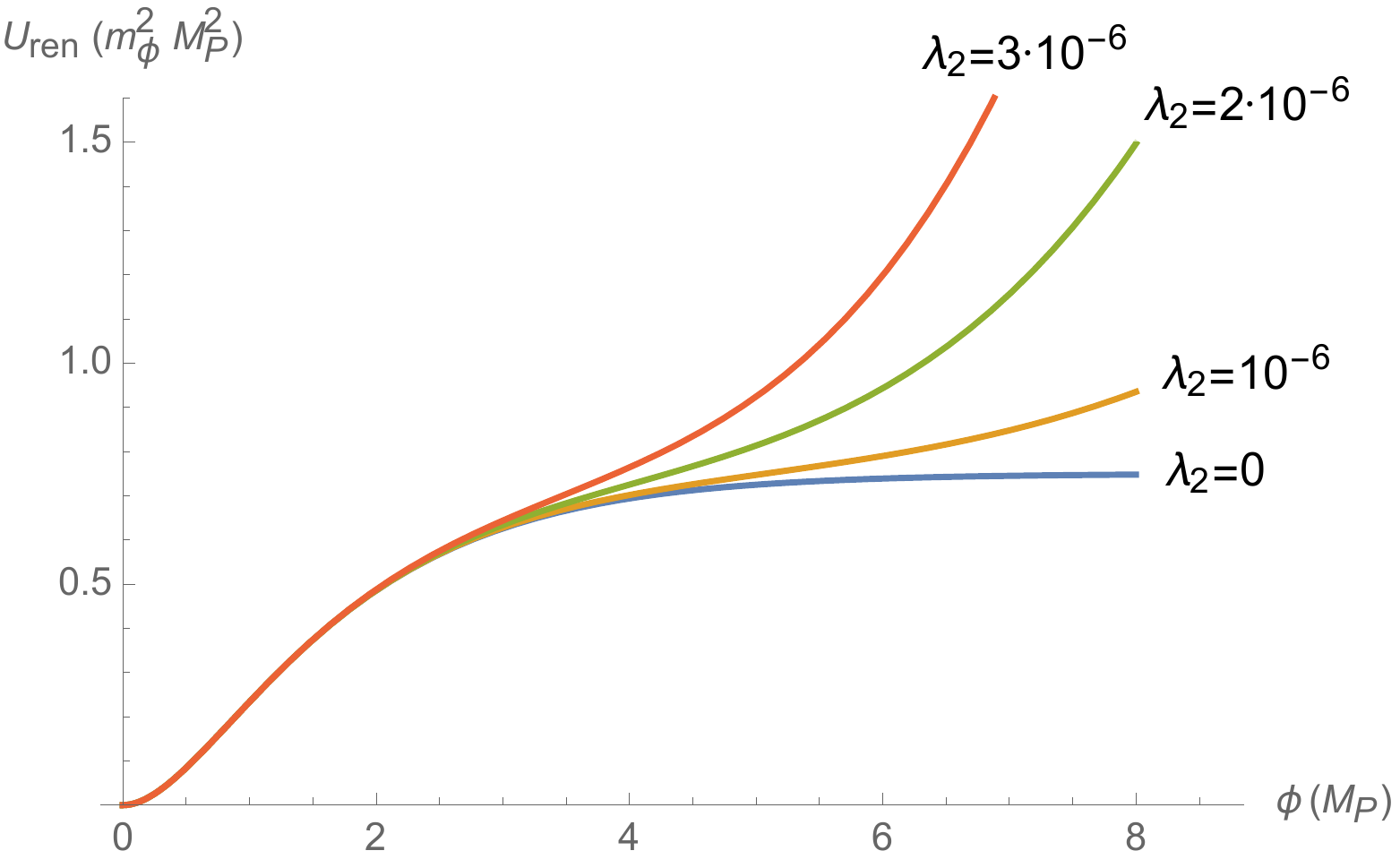}
\end{center}
\caption{The renormalized effective potential in equation (\ref{eq:Uren}) for different values of $\lambda_2$ in the Starobinsky model of inflation. The ratio  $m_\phi/M_P$ is fixed by the scalar primordial amplitude. The large values of $\lambda_2$ necessary for an efficient reheating spoil the flatness of the potential. 
\label{fig:Uren} } 
\end{figure}

With the given form of the potential (\ref{eq:Starobinsky}), in order for the universe to inflate more than sixty e-folds, the initial value of $\phi$ has to be larger than about $5.5 \, M_P$. The amplitude of the primordial perturbations then fixes the inflaton mass to  
\begin{equation}\label{eq:mphi primordial}
	m_\phi= 1.2 \times 10^{-5} M_P,
\end{equation}
and the validity of perturbation theory  demands that the coupling to matter be small, $\lambda_2\ll 1$. Actually,  if $\lambda_2$ is not small enough,  radiative corrections prevent us from approximating the actual effective potential by  (\ref{eq:Starobinsky}). Equation  (\ref{eq:Uren}) in particular implies that, in order for the relative difference between $V$ and $U_\mathrm{ren}$  to remain  less than $5\%$ at $\phi\leq 5.5\, M_P$,  the coupling constant  needs to satisfy   $\lambda_2\lesssim 10^{-6}$ (see figure \ref{fig:Uren}.) This condition constrains the parameter $q_0$ to the range 
\begin{equation}\label{eq:q0 range}
	q_0\lesssim 5\times10^3,
\end{equation}
which leaves ample room for parametric resonance. Because we keep the mass $m_\phi$ fixed, we shall quote the values of $q_0$, rather than those of the proportional $\lambda_2$ in what follows.

Yet, as we argued in section \ref{sec:Backreaction Numerical}, the parameter $q_0$ cannot be too small either, because otherwise no successful reheating takes place. By comparing the numerically determined background energy density of the inflaton $\bar\rho_\phi$ to that of matter $\braket{\rho_\chi}_\mathrm{ren}$, we observe that  in order for the universe to become dominated by radiation after  the end of inflation  it is necessary (but not sufficient) that the resonance parameter obey $q_0 \gtrsim 10^4$, which  is marginally incompatible with equation (\ref{eq:q0 range}).

The previous arguments illustrate the power of our analysis. Typically, constraints on inflationary models only concern the form of the effective  potential, since this is what directly determines the spectra of primordial perturbations. Demanding a successful reheating  places additional  conditions on the couplings to matter, which, when combined with limits on the size of the radiative corrections,  can be quite restrictive. In the example of the Starobinsky model, with our chosen couplings to scalar matter and within our approximations, it appears that one cannot have a successful reheating and the prescribed form of the effective potential at the same time. This does not necessarily mean that the Starobinsky potential is inviable, since other couplings to matter are possible, but it does point out that there is more to an inflationary model than just the form of the scalar potential. 

\subsection{Gravitational Wave Spectra}

\begin{figure}
\subfloat[Predicted Strain] 
{
\includegraphics[width=7.5cm]{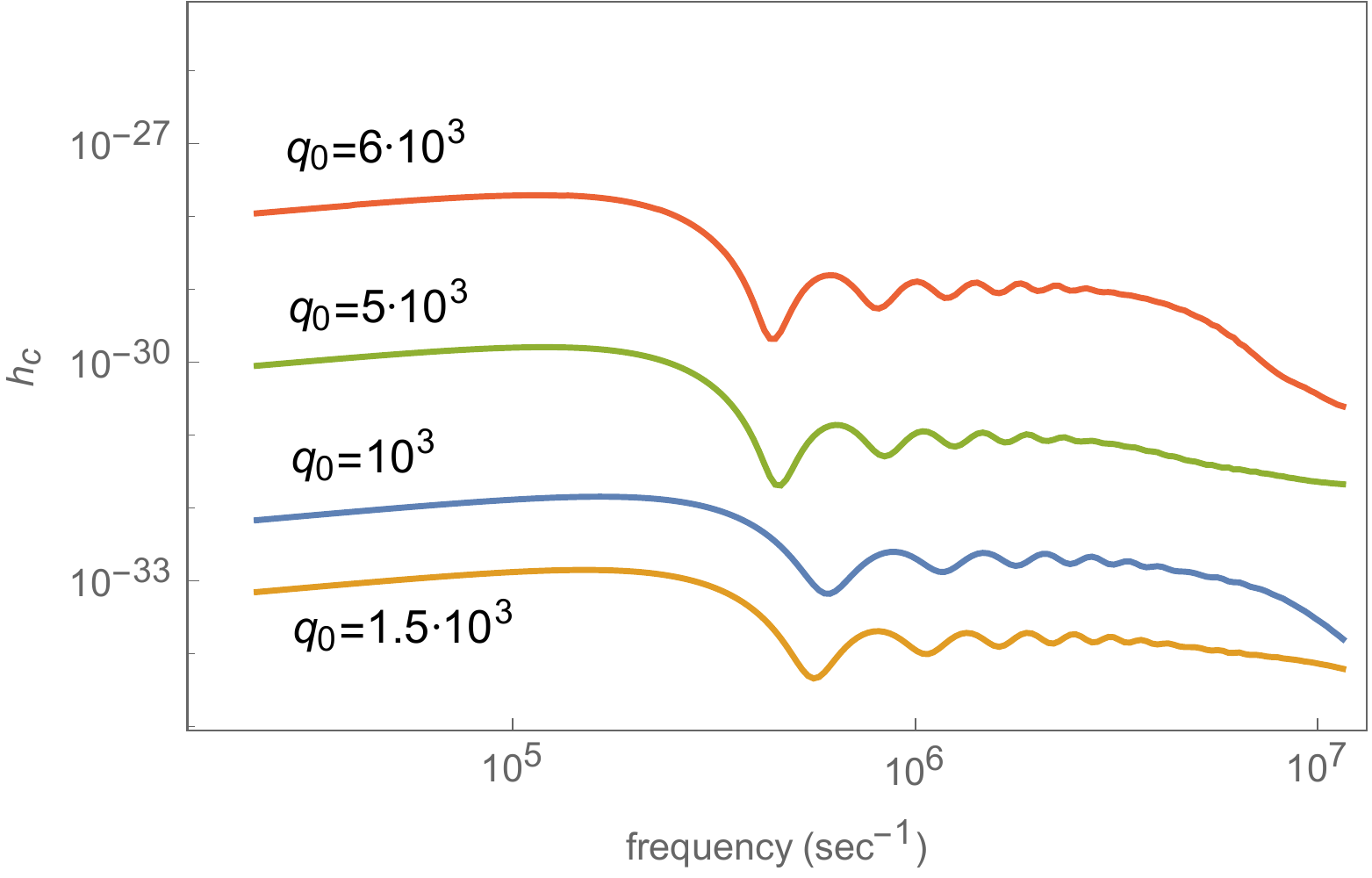}
}
\subfloat[Detection Prospects]
{
\includegraphics[width=7.5cm]{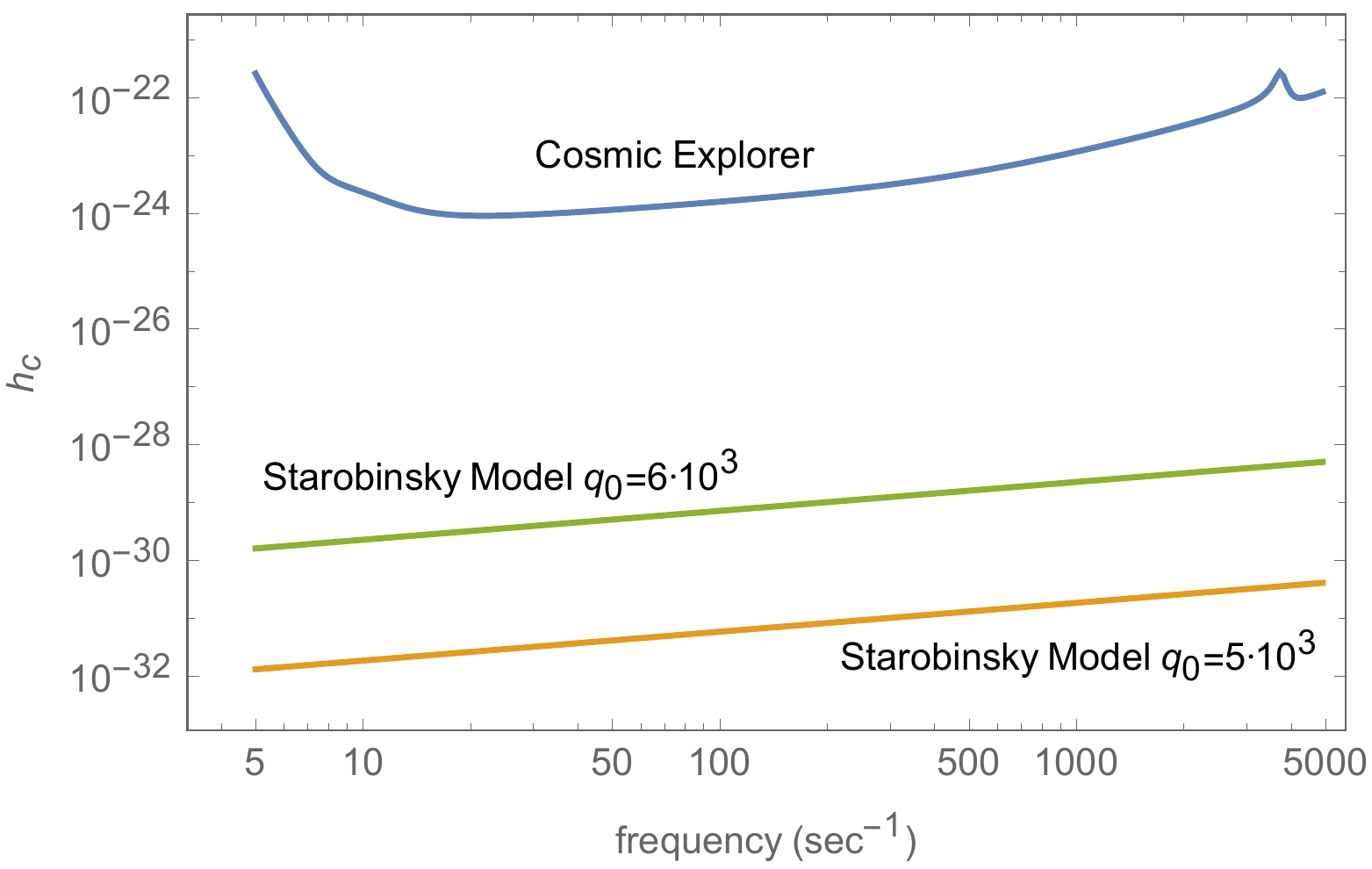}
}
\caption{Panel (a) Expected strain of the gravitational waves produced in the Starobinsky model. Panel (b) Cosmic Explorer sensitivity compared with the  predicted strain at low frequencies.} 
\label{fig:strain}
\end{figure}

We plot  present-day spectra of the gravitational waves  produced during preheating in the Starobinsky model in panel (a) of figure \ref{fig:strain}.  In order to allow a direct   comparison with detector sensitivities, we actually display the expected characteristic  strain  of the  signal,  $h_c(f)$, which  is related to the spectral density  $\Omega^0_\mathrm{GW}(f)$   defined in reference \cite{Armendariz-Picon:2019csc} by  
\begin{equation}
	h^2_c(f)=\frac{3H_0^2}{2\pi^2 f^2}\Omega^0_\mathrm{GW}(f).
\end{equation}
Here,  $H_0$ is today's Hubble constant and $f$ is the gravitational wave frequency \cite{Moore:2014lga}.  The frequency $f$ of a gravitational wave  is simply proportional to its comoving momentum $p$, see also reference \cite{Armendariz-Picon:2019csc} for the exact relation. We keep $m_\phi$ fixed at the value quoted in equation (\ref{eq:mphi primordial}), and vary $q_0$  in the range compatible with the form of the effective potential (\ref{eq:q0 range}).  As mentioned in the previous subsection, the latter is marginally incompatible with a successful reheating process. Because the final amplitude of the gravitational waves today depends on how long it takes for the universe to become radiation-dominated after the end of inflation, our predictions should not be taken too literally.

As seen in figure \ref{fig:strain}, the predicted gravitational wave signal is not a monotonic function of $q_0$.  Although this is not shown, the impact of backreaction sets in relatively abruptly,  around $q_0\sim 2\cdot 10^3$. This happens to be the value of $q_0$ for which $\braket{\chi^2}_\mathrm{ren}/m_\phi^2$ surpasses $\lambda_2^{-1}$.  For values of $q_0$ in the range $2\cdot 10^3\leq q_0\lesssim 5\cdot 10^3$, the  spectrum is relatively insensitive to the precise value of $q_0$, particularly in the low frequency asymptote, although the signal does slightly grow with $q_0$ in that regime. The largest strain in the Starobinsky model compatible with (\ref{eq:q0 range}) is thus achieved at $q_0=5\cdot 10^3$.  For illustration we also plot one  spectrum beyond this value. Unfortunately, though, our somewhat rudimentary numerical implementation does not allow us to probe much higher values of $q_0$, as the gravitational spectra become noisy and unreliable. Nevertheless, from the dependence of $\braket{\chi^2}_\mathrm{ren}$ on $q_0$,  we do not  expect the signal  to grow beyond $q_0\sim 10^4$. 

Regrettably, the gravitational wave spectra peak at  frequencies beyond the sensitivity of current and near-future detectors, which extends at most to frequencies of about $10^4$ Hz. In order to obtain an estimate of the gravitational wave signal around those frequencies, we note that the results of reference \cite{Armendariz-Picon:2019csc} imply that the gravitational wave spectrum is proportional to $|u_p(t_f)|^2$, where  $u_p(t_f)$ is the mode function of a gravitational wave of momentum $p$ at the end of reheating.\footnote{To see this, consider equation (4.26) in \cite{Armendariz-Picon:2019csc}. On super-horizon scales the mode functions $u_p(\bar{t}_1)$ are essentially constant during preheating, and can be taken out of the integral. The spectrum is thus proportional to $|u(t_f)|^2$, as claimed.} At low frequencies  we expect a nearly scale invariant spectrum from inflation, so $|u_p(t_f)|^2\propto p^{n_T-3}$, where $n_T\approx 0$ is the tensor spectral index.  This translates into a relatively flat characteristic strain proportional to $f^{1/2}$, which happens to fit the slow rise of $h_c$ seen at $f\leq 10^5$ Hz in panel (a) of figure \ref{fig:strain}.

In panel (b) of  figure \ref{fig:strain} we plot the low-frequency tail of the expected gravitational wave signal, in conjunction with the sensitivity curve of  a next generation gravitational wave detector, the ``Cosmic Explorer" \cite{Evans:2016mbw} (the sensitivity data was downloaded from the url listed in \cite{Moore:2014lga}.)  In the present case, even if we push the boundary a bit by considering $q_0=6\cdot 10^6$,  sensitivity and signal are separated by several orders of magnitude, making a detection in the near future highly unlikely.

\section{Summary and Conclusions}
 
Motivated by our previous analysis  of gravitational wave production during preheating,  we have studied how to incorporate backreaction into the motion of a homogeneous scalar field coupled to matter in an expanding universe. Backreaction affects the motion of the scalar not only through its direct couplings to matter, but  also indirectly through the change in the expansion history linked to the energy-momentum tensor of matter.

In order to derive the effective equations of motion both for the inflaton and the scale factor, it is convenient to   demand  the vanishing of the  expectation of the inflaton and metric fluctuations. This immediately leads to physically transparent quantum-corrected equations  and avoids the cumbersome quantum effective action in the $in$-$in$ formalism.  In addition the formalism does not rely on the particle concept and circumvents many of the ambiguities and shortcomings of the latter. 

All the expectation values that appear in the quantum corrected equations of motion are ultraviolet divergent, and thus require regularization and renormalization. In this context, Pauli-Villars regularization proves to be extremely useful. It allows us to regularize the theory with a physically transparent procedure, while preserving diffeomorphism invariance, within a setting  that is  readily amenable to numerical integration of the ensuing mode integrals.  Quantum corrections to the equations of motion can be expanded according to the number of time derivatives acting on the background quantities. Only the first few terms in this expansion require renormalization, but to fully take into account quantum corrections to its motion one needs to include all the terms in the series.  

We have applied this formalism to incorporate backreaction into the production of gravitational waves during preheating. This basically completes our program of making rigorous predictions for the expected gravitational wave signal  from first principles. The extent of the agreement between the standard numerical methods and ours remains to be assessed. In our opinion, the manifest differences between the two approaches  calls into question the accuracy of the standard numerical predictions. We leave it to the proponents of the standard numerical approaches to properly evaluate and  justify their methods. 

As an illustration of our approach, we have chosen to focus on  gravitational wave production in the scalar version of the Starobinsky model, when the inflaton couples to matter through a quartic coupling.  In order for reheating to be efficient, it is necessary for the coupling strength to be large, but the latter is incompatible with the assumed form of the  effective potential. In particular, it appears that within this model it is not possible to successfully reheat the universe while keeping the effective potential sufficiently flat.  We expect this tension between the needs for  efficient reheating and controlled radiative corrections to persist in any other inflationary model.  In this particular case, the strong suppression of the signal with the fourth power of the inflaton mass renders gravitational waves undetectable in the near future, in spite of the strong enhancement by parametric resonance.   All these observations underscore that, as more phenomenological predictions become feasible within the inflation  paradigm, there is more to inflationary model-building that just the form of the  effective potential.

\appendix

\section{Pressure Renormalization}
\label{sec:Pressure Renormalization}

Although we have  mostly concentrated  on the  Friedman equation to asses the impact of backreaction on the evolution of the scale factor, for some purposes it may be also  useful to study the spatial components of  the semiclassical Einstein equation
\begin{equation}\label{eq:Einsteinij semi}
	\frac{\ddot{a}}{a^3}=\frac{1}{6M_P^2}\left(\braket{\rho}-3\braket{p}\right),
\end{equation}
where  the  pressure of matter is  
\begin{equation}\label{eq:p def}
	p_\chi \delta^{ij} \equiv a^2 \langle T^{ij}_{(\chi)}  \rangle.
\end{equation}
This expectation is again divergent and requires renormalization. To determine its renormalized value we proceed as with the energy density. We begin by expanding in the number of time derivatives,
\begin{subequations}\label{eq:p expansion}
\begin{align}
	3p_\chi^{(0)}&= \frac{1}{8 (2\pi^2)} \sum_i \sigma_i \left[
	\frac{\Lambda^4}{a^4}-\frac{\Lambda^2 \kappa_i}{a^2}-\frac{\kappa_i^2}{4}\left(\frac{7}{2}-3\log x_i\right)\right], \label{eq:p0}
	\\
	\begin{split}
	3p_\chi^{(2)}&= \frac{1}{8 (2\pi^2)} \sum_i \sigma_i \bigg[
	\frac{3\Lambda^2\mathcal{H}^2}{a^4}
	-\frac{2\Lambda^2 \ddot{a}}{a^5}
	-\frac{\kappa_i\mathcal{H}^2}{2a^2}\left(\frac{2}{3}-\log x_i)\right)
	+\frac{\mathcal{H}\dot{\kappa}_i}{2a^2}\left(\frac{17}{3}-\log x_i\right)\\
	&\quad +\frac{\kappa_i \ddot{a}}{a^3}\left(\frac{8}{3}-\log x_i\right)	
	+\frac{\ddot{\kappa}_i}{2a^2}\left(\frac{5}{3}-\log x_i\right)+ \frac{5\dot{\kappa}_i^2}{8a^2 \kappa_i}\bigg], \label{eq:p2}
	\end{split}
	\\
	\begin{split}
	3p_\chi^{(4)}&=\frac{1}{8 (2\pi^2)} \sum_i \sigma_i \bigg[
	-\frac{\mathcal{H}^4}{12a^4}+\frac{\mathcal{H}^2 \ddot{a}}{a^5}\left(1+4\log x_i\right)
	+\frac{\ddot{a}^2}{4a^6}\left(\frac{1}{3}-5\log x_i\right) \\
	&\quad +\frac{\mathcal{H}\dddot{a}}{2a^5}\left(\frac{7}{3}-5\log x_i\right)
	-\frac{a^{(4)}}{2a^5}\left(\frac{19}{15}-\log x_i\right)\bigg], \label{eq:p4}
	\end{split}
\end{align}
\end{subequations}
where in equation (\ref{eq:p4}) we have omitted  terms that vanish in the limit $M_i\to\infty$.  As before, the cutoff dependent terms cancel because of equations (\ref{eq:sigma cond}) and (\ref{eq:add sigma cond}),  but a dependence on the regulator masses remains. The latter is canceled by the counterterms, which effectively contribute to the pressure at different orders in the derivative expansion,
\begin{subequations}\label{eq:p ct}
\begin{align}
	3p^{(0)}_{\mathrm{ct}}&=-3\delta\Lambda-3\delta d_1\bar\lambda-3\delta d_2 \bar{\lambda}^2,\\
	3p^{(2)}_{\mathrm{ct}}&=3\delta M_P^2\left[\frac{2\ddot{a}}{a^3}-\frac{\mathcal{H}^2}{a^2}\right]
		+6\delta \xi_f\left[\frac{\mathcal{H}^2 \bar{\lambda}}{a^2}-\frac{\mathcal{H}\dot{\bar\lambda}}{a^2}-
		\frac{2\ddot{a}\bar\lambda}{a^3}-\frac{\ddot{\bar\lambda}}{a^2}\right],\\
	3p^{(4)}_{\mathrm{ct}}&=\left(3\delta c_{(1)}+\delta c_{(2)}\right)\left[\frac{192 \mathcal{H}^2 \ddot{a}}{a^5}-\frac{60 \ddot{a}^2}{a^6}-\frac{120\mathcal{H}\dddot{a}}{a^5}+\frac{24 a^{(4)}}{a^5}\right].
\end{align}
\end{subequations}
Remarkably, the very same  counterterms that eliminate the dependence of the energy density on the regulator masses cancel that of  the pressure.  Combining both sets of equations (\ref{eq:p expansion})  and (\ref{eq:p ct}) we arrive at the renormalized pressure
\begin{subequations}\label{eq:p ren}
\begin{equation}
\begin{split}
3\braket{p_\chi}_\mathrm{ren}&=\frac{1}{2\pi^2}\bigg\{
\frac{1}{2a^2} \int^\Lambda dk \, k^2 \left[3\dot{w}_k \dot{w}_k^*-(k^2+3a^2 \kappa_0)w_k w_k^* \right] 
\\
&-\frac{\Lambda^4}{8a^4}+\frac{\Lambda^2\kappa_0}{8a^2}-\frac{3\kappa_0^2}{32}\log \frac{4\Lambda^2}{a^2\kappa_\mu}+g_0
\\
&-\frac{3\Lambda^2 \mathcal{H}^2}{8a^4}+\frac{\Lambda^2\ddot{a}}{4a^5}
-\frac{\kappa_0}{16}\left(\frac{\mathcal{H}^2}{a^2}-\frac{2\ddot a}{a^3}\right)\log \frac{4\Lambda^2}{a^2\kappa_\mu}
+\frac{1}{16}\left(\frac{\mathcal{H}\dot{\bar\lambda}}{a^2}+\frac{\ddot{\bar\lambda}}{a^2}\right)\log \frac{4\Lambda^2}{a^2\kappa_\mu}
+g_2
\\
&-\frac{1}{8}\left(\frac{4\mathcal{H}^2\ddot{a}}{a^5}-\frac{5}{4}\frac{\ddot{a}^2}{a^6}-\frac{5}{2}\frac{\mathcal{H}\dddot{a}}{a^5}+\frac{1}{2}\frac{a^{(4)}}{a^5}\right)\log  \frac{4\Lambda^2}{a^2\kappa_\mu}+g_4
\bigg\},
\end{split}
\end{equation}
where the cutoff independent pieces are 
\begin{align}
 g_0 &=\frac{7\kappa_0^2}{64}+\frac{3}{32}\left(M_0^2\bar\lambda+\frac{3}{2}\bar\lambda^2\right)-3(\delta\Lambda^f+\delta d_1^f \bar{\lambda}+\delta d_2^f \bar\lambda^2), 
 \\ 
 \begin{split}
 g_2 &= \frac{(2\kappa_0+3\bar\lambda)\mathcal{H}^2}{48}-\frac{(8\kappa_0+3\bar\lambda) \ddot{a}}{24a^3}-\frac{17 \mathcal{H}\dot{\bar\lambda}}{48 a^2}-\frac{5 \ddot{\bar\lambda}}{48a^2}\\
 &\quad+\left[3 (\delta M_P^2)^f-6 \delta\xi^f \bar\lambda\right]\left(\frac{2\ddot{a}}{a^3}-\frac{\mathcal{H}^2}{a^2} \right)
-6\delta\xi^f \left(\frac{\mathcal{H}\dot{\bar\lambda}}{a^2}+\frac{\ddot{\bar\lambda}}{a^2}\right) ,
\end{split}
\\
\begin{split}
 g_4&= \frac{1}{8}\left(\frac{\mathcal{H}^4}{12a^4}-\frac{\mathcal{H}^2 \ddot{a}}{a^5}-\frac{\ddot{a}^2}{12a^6}-\frac{7 \mathcal{H}\dddot{a}}{6a^5}+\frac{19 a^{(4)}}{30a^5}\right)
  +48\delta c^f \left(\frac{4\mathcal{H}^2 \ddot{a}}{a^5}-\frac{5 \ddot{a}^2}{4a^6}-\frac{5\mathcal{H}\dddot{a}}{2a^5}+\frac{a^{(4)}}{2a^5}  \right).
 \end{split}
\end{align}
Note that some of the terms  contain three of more derivatives of the scale factor and two derivatives of the scalar. In order to preserve the order of the original differential equation (\ref{eq:Einsteinij}), we thus proceed to reduce the order of these corrections, as in the case of the energy density. Using  the lowest order equation (\ref{eq:Einsteinij}) and $\ddot{\bar\phi}+2\mathcal{H}\dot{\bar\phi}+a^2 \bar V'=0$ we conclude that the four derivative terms in equations (\ref{eq:p ren}) should be replaced by 
\begin{equation}
\begin{split}
&\left(48 \delta c^f-\frac{1}{8}\log \frac{4\Lambda^2}{a^2 \kappa_mu}\right)
\bigg[
-\frac{\mathcal{H}^2(35 \dot{\bar\phi}^2+4a^2 \bar V)}{12 M_P^2 a^4}
-\frac{3\mathcal{H} \dot{\bar V}}{2M_P^2 a^2} 
+\frac{\bar{V}'' \dot{\bar\phi}^2-a^2\bar{V}'^2}{2M_P^2 a^2}
\\
&-\frac{11\dot{\bar\phi}^4-40a^2\bar{V} \dot{\bar\phi}^2-16a^4 \bar V^2 }{144 M_P^4 a^4}
\bigg]
+\frac{1}{8}\bigg[
\frac{\mathcal{H}^4}{12a^4}
-\frac{\mathcal{H}^2 (101\dot{\bar\phi}^2+28 a^2 V)}{60 M_P^2 a^4}
\\
&+\frac{\mathcal{H} \dot{\bar V}}{10M_P^2 a^2}
-\frac{19 (a^2\bar V'^2 - \bar{V}'' \dot{\bar\phi}^2) }{30 M_P^2 a^2}
-\frac{119 \dot{\bar\phi}^4-40a^2 \bar V \dot{\bar\phi}^2-1744a^4 \bar V^2 }{2160 M_P^4 a^4}
 \bigg].
\end{split} 
\end{equation}
\end{subequations}

\section{Covariant Conservation}
\label{sec:Covariant Conservation}

In our analysis of backreaction we have concentrated on the time-time component of the Einstein equations, because  diffeomorphism invariance relates the former to their spatial components.  Indeed, both in the $in$-$in$ and $in$-$out$ formalisms the invariance of the   action under diffeomorphisms, along with the validity of the equations of motion inside an expectation,  implies the  conservation  of the energy-momentum tensor,
\begin{equation}\label{eq:cov cons}
	\nabla_\mu \langle  T^{\mu\nu}\rangle=0.
\end{equation}
 Because the Einstein tensor obeys the Bianchi identity $\nabla_\mu G^{\mu\nu}=0$,  equation (\ref{eq:cov cons}) is necessary for the self-consistency of the semiclassical Einstein equations (\ref{eq:Einstein eqs}), and  it also allows one to derive the spatial components of Einstein's equations from the time derivative of the time components.  Our goal here is to check whether the  expectation of the energy-momentum tensor actually obeys (\ref{eq:cov cons}), which allows us to test  whether our regularization scheme preserves diffeomorphism invariance.  We shall do so in the regularized theory alone, with finite and arbitrary regulator masses, since the contribution of the counterterms is automatically covariantly conserved, again by diffeomorphism invariance.   In addition we shall  see how equation (\ref{eq:cov cons}) is related to the quantum-corrected field equations of motion.

It is again convenient to consider the  split of the energy-momentum tensor in equation (\ref{eq:split}). Because of the isometries of the cosmological background, the only non-trivial conservation equation involves the $\nu=0$ component,
\begin{align}
a^4\nabla_\mu \left\langle T^{\mu 0}_{(\phi)}\right\rangle&=\left[\ddot{\bar\phi}+2\mathcal{H} \dot{\bar\phi}+a^2 \bar{V}'\right]\dot{\bar\phi},
\\
a^2 \nabla_\mu \left\langle T^{\mu 0}_{(\chi)}\right\rangle&=\dot\rho_\chi+3\mathcal{H} ( \rho_\chi+p_\chi),
\end{align}
where the energy density and pressure of matter are respectively defined in equations (\ref{eq:Friedman}) and (\ref{eq:p def}).  Although our notation appears to imply that there is only a single matter field $\chi$, what we really mean is a sum over all fields $\chi_i$, as in the main text.  The covariant conservation of the total energy-momentum tensor reduces to the condition
\begin{equation}\label{eq:final cov cons}
	\left[\ddot{\bar\phi}+2\mathcal{H} \dot{\bar\phi}+a^2 \bar V'\right]\dot{\bar\phi}
	+\big[\dot\rho_\chi+3\mathcal{H} ( \rho_\chi+p_\chi)\big] a^2=0.
\end{equation}
Incidentally, because $\partial_j\braket{\partial^j \chi\,  \dot{\chi}}=0$ by symmetry, the previous equation can be  also cast   as
\begin{equation}
\left[\ddot{\bar\phi}+2\mathcal{H} \dot{\bar\phi}+a^2 \bar V' +a^2 \frac{\bar\lambda'}{2}\braket{\chi^2}\right]\dot{\bar\phi}
	+\Braket{(-\Box \chi + \kappa_0 \chi)\dot{\chi}}a^2=0,
\end{equation}
which illustrates that the expectation of the energy-momentum  is conserved if the appropriate equations of motion hold. In particular, the term multiplying $\dot{\bar{\phi}}$ is the quantum-corrected inflaton equation of motion, and the expectation multiplying $a^2$ vanishes because in the interaction picture the matter mode functions satisfy the  mode equation  (\ref{eq:mode equation}). 

Returning to equation  (\ref{eq:final cov cons}), we can then  establish that the energy momentum tensor is covariantly conserved   if we verify  that
\begin{equation}\label{eq:equivalence}
 	\dot{\rho}_\chi+3\mathcal{H} (\rho_\chi+p_\chi)=\frac{\bar\lambda'}{2}\dot{\bar\phi}\braket{\chi^2},
\end{equation}
for the latter then yields the regularized quantum corrected inflaton equation of motion. The non-zero term on the right hand side of equation (\ref{eq:equivalence}) simply arises because the inflaton and matter are coupled. Whenever $\bar{\lambda}$ increases, energy is transferred away from the inflaton to matter, and vice versa.     It is not difficult to prove  equation (\ref{eq:equivalence}) by expressing $\rho_\chi$ and $p_\chi$ in terms of the mode functions of the different fields, $w^i_k$, and using the analogue of equation (\ref{eq:mode equation}) to eliminate $\ddot{w}^i_k$. All one needs to assume is that the mode integrals are finite because of the contribution of the regulator fields.  To clear any lingering doubts, we shall  also verify equation (\ref{eq:equivalence}) by expanding again in the number of time derivatives.  At one derivative  equation (\ref{eq:equivalence})  holds because equations   (\ref{eq:chisq0}), (\ref{eq:T000})  and (\ref{eq:p0})  imply that 
\begin{equation}
\dot\rho_\chi^{(0)}+3\mathcal{H} (\rho^{(0)}_\chi+p^{(0)}_\chi)-\frac{\bar{\lambda}'}{2}\dot{\bar\phi} \braket{\chi^2}^{(0)}=\frac{1}{32(2\pi^2)}\sum_i \sigma_i \kappa_i \dot\kappa_i,
\end{equation}
and the right hand vanishes because of equations (\ref{eq:kappai}) and (\ref{eq:sigma cond}). Similarly, combining equations (\ref{eq:chisq2}),  (\ref{eq:T002}) and (\ref{eq:p2})   we find
\begin{equation}
	\dot \rho_\chi^{(2)}+3\mathcal{H} (\rho^{(2)}_\chi+p^{(2)}_\chi)-\frac{\bar{\lambda}'}{2}\dot{\bar\phi}\braket{\chi^2}^{(2)}=-\frac{1}{8(2\pi^2)}\sum_i \sigma_i   \frac{\mathcal{H}^2\dot\kappa_i}{2a^2},
\end{equation}
which vanishes for the same reason. 
 
 \section{Equations of Motion}
 \label{sec:Equations of Motion}
 
In order to  numerically solve the equations of motion that determine the evolution of the inflaton and the scale factor, it shall prove convenient to switch to cosmic time as integration variable. It is also useful to formulate all the equation in terms of dimensionless variables, which allows us  to study which dimensionless parameters control the different limits of the evolution. 
 
 Let us assume for simplicity that the scalar field potential and the coupling function have the form
 \begin{equation}
 	V(\phi)=M_P^2 m_\phi^2\,  V_0\left(\frac{\phi}{M_P}\right), 
	\quad
	\lambda=M_P^2\,  \lambda_0\!\left(\frac{\phi}{M_P}\right),
 \end{equation}
 where, without loss of generality, $V'(0)=0$ and $V''_0(0)=1$. The first condition on the potential  implies that the minimum of the potential is at $\phi=0$, while the second establishes the effective mass of the field at the minimum equals $m_\phi^2$. Introducing the dimensionless time variable $x=m_\phi \tau$, where $\tau$ is cosmic time, and the dimensionless field $\phi_P=\phi/M_P$, the quantum-corrected inflaton equation of motion (\ref{eq:quantum eom})  becomes
 \begin{equation}\label{eq:numerical eom}
\frac{d^2 \phi_P}{dx^2}+3 H_x \frac{d\phi_P}{dx}+\Braket{\frac{\partial V_0}{\partial \phi_P}}_\mathrm{ren}=0,
\quad H_x \equiv\frac{1}{a}\frac{da}{dx}.
\end{equation}
The Hubble constant in units of $m_\phi$ then obeys the modified Friedman equation
\begin{equation}\label{eq:num Friedman}
	H_x^2=\frac{1}{3}\left[\frac{1}{2}\left(\frac{d\phi_P}{dx}\right)^2+V_0(\phi_P)+\frac{m_\phi^2}{M_P^2}\frac{\braket{\rho_\chi}_\mathrm{ren}}{m_\phi^4}\right],
\end{equation}
where the energy density of matter is, from equation (\ref{eq:T00 final}),
\begin{equation}\label{eq:rho code}
\frac{\braket{\rho_\chi}_\mathrm{ren}}{m_\phi^4}=
\frac{1}{2\pi^2}\left\{\frac{1}{2}\int_0^{\tilde \Lambda} d\tilde k\,  \tilde{k}^2 
\left[\frac{d\tilde w_{\tilde k}^*}{dx} \frac{d\tilde w_{\tilde k}}{dx} +
\left(\tilde \kappa_0+\frac{\tilde{k}^2}{a^2}\right)\tilde w_{\tilde k}^* \tilde w_{\tilde k} \right] 
-\frac{\tilde \Lambda^4}{8a^4}+\cdots\right\}.
\end{equation}
In the previous equation a tilde denotes the dimensionless variable obtained by multiplication with the appropriate power of  $m_\phi$, say,
\begin{equation}
	\tilde{\Lambda}\equiv \frac{\Lambda}{m_\phi},  \quad
	\tilde{k}\equiv \frac{k}{m_\phi},  \quad
	\tilde{w}_{\tilde{k}}=m_\phi^{1/2} w_{\tilde{k}}, \quad
	\tilde\kappa_0=\frac{M_0^2 +M_P^2\, \lambda_0(\phi_P)}{m_\phi^2}.
\end{equation}
From equation (\ref{eq:mode equation}), the rescaled mode functions $\tilde{w}_{\tilde k}$ obey the differential equation
\begin{equation}\label{eq:num mode}
	\frac{d^2 \tilde{w}_{\tilde{k}}}{dx^2}+3 H_x \frac{d\tilde{w}_{\tilde{k}}}{dx}+\left(\frac{\tilde{k}^2}{a^2}+\tilde{\kappa}_0\right)\tilde{w}_{\tilde{k}}=0. 
\end{equation}
with initial conditions that can be fixed by matching  the  adiabatic vacuum 
\begin{equation}
\tilde{w}_{\tilde{k}}\approx \frac{\exp\left(-i\int^x  \tilde{W}_{\tilde{k}} \, dx_1\right)}{a^{3/2}\sqrt{2 \tilde{W}_{\tilde{k}}}} 
, \quad
\tilde{W}_{\tilde{k}}\approx \left(\frac{\tilde{k}^2}{a^2}+\tilde \kappa_0\right)^{1/2}+\cdots.
\end{equation}
At early times the matching is possible for all modes of interest only if the universe underwent a sufficiently long period of inflation. It is often simpler to work with rescaled mode functions $v_k\equiv w_k a^{3/2}$, because the friction term proportional to $-3H_x$ drops out of the  mode equation.  But  in that case the dispersion relation of the $v_k$ contains second time derivatives of the scale factor, which makes these variables less convenient overall.

Finally, the renormalized and dimensionless driving term in equation (\ref{eq:numerical eom}) reads
\begin{equation}
	\Braket{\frac{\partial V_0}{\partial \phi_P}}_\mathrm{ren}=
	\bar{V}_0'+
	\frac{\bar{\lambda}_0'}{2}\frac{\braket{\chi^2}_\mathrm{ren}}{m_\phi^2},
\end{equation}
where the expectation on the right hand side is, from equation (\ref{eq:ren driving}), 
\begin{equation}
\begin{split}
\frac{\braket{\chi^2}_\mathrm{ren}}{m_\phi^2}=
\frac{1}{2\pi^2}\Bigg[
\int_0^{\tilde\Lambda} \!\!\! d\tilde k \, \tilde{k}^2 \,|\tilde w_{\tilde k}|^2
-\frac{1}{4}\frac{\tilde \Lambda^2}{a^2}&-\frac{\tilde \kappa_0}{8}\left(1-\log x_\mu\right)
	-\frac{\tilde R}{24}\left(-\frac{5}{6}+\frac{1}{2}\log x_\mu \right)\\
	&+2\frac{\delta d_1^f}{m_\phi^2}+\frac{M_P^2}{m_\phi^2}\left(4\delta d_2^f-\frac{1}{16}\right)\bar{\lambda}_0+2\delta \xi^f \tilde{R}\Bigg]. \nonumber
\end{split}
\end{equation}
Note that $x_\mu=4\tilde{\Lambda}^2/(a^2 \tilde{\kappa}_\mu)$  is already dimensionless.

We have thus managed to express all equations of motion in terms of dimensionless quantities. The previous equations  indicate that $m_\phi^2/M_P^2$ is the parameter that determines the importance of backreaction on cosmic expansion,   $\tilde{\kappa}_0$ is the function responsible for parametric resonance, and  $\bar{\lambda}_0'$ is the function that characterizes the importance of backreaction on the inflaton motion. 

Yet there is an additional modification that needs to be made. The standard  mode equation for the tensor modes  used in reference \cite{Armendariz-Picon:2019csc} assumes that the background satisfies the classical Einstein equations.  Because our  background   satisfies the semiclassical equations  (\ref{eq:Einstein eqs}) instead, the evolution equation for the gravitational wave modes is
\begin{equation}
	\frac{d^2{h}_{ij}}{dx^2}+3H_x \frac{d\dot{h}_{ij}}{dx}+\left[\frac{\tilde{p}^2}{a^2}-2\frac{m_\phi^2}{M_P^2}\frac{\braket{p_\chi}}{m_\phi^4}\right]h_{ij}=0,
\end{equation}
where $\braket{p_\chi}$ is the renormalized matter pressure in equation (\ref{eq:p ren}), and $\tilde{p}$ the comoving momentum of the wave in units of $m_\phi$. In the limit of no backreaction, $m_\phi^2/M_P^2\to 0$, this equation clearly reduces to the standard result.
 
For our numerical implementation we need to choose a specific model. We can always approximate the scalar potential around the minimum by a quadratic function, and we shall assume  assume that the coupling function is quadratic too,
\begin{equation}\label{eq:illustration}
	V_0=\frac{1}{2} \phi_P^2, \quad \lambda_0=\lambda_2 \phi_P^2. 
\end{equation}
To avoid the introduction of further dimensionless parameters, we shall also choose $M_0=0$.  In that case, $\tilde{\kappa}_0=\lambda_2 (M_P^2/m_\phi^2)\phi_P^2$, so  $q_0$ in equation (\ref{eq:q0}) controls the effectiveness of parametric resonance, and $\bar{\lambda}_0'=2\lambda_2\phi_P$ determines the impact of backreaction on the motion of the inflaton. 
 
 \section{Boundary Terms}
 \label{sec:Boundary Terms}
 
 Thus far we have ignored the different boundary terms  we have  encountered throughout our analysis. In the $in$-$out$ formalism this is well-justified as the $i\epsilon$ prescription eliminates the contributions of the fields in the infinite asymptotic past and future. But in the $in$-$in$ formalism, the evaluation of expectation values at a finite time $t$ effectively introduces a boundary  in the spacetime at that time, $\partial\mathcal{M}$, at which the fields do not vanish and interactions are still effective.  
 
 Such boundary terms are relevant because they may contribute to the expectation of the field operators in some cases. Indeed, as argued in \cite{Armendariz-Picon:2019csc},    when the  interaction Lagrangian  contains a total derivative $L_I=-dB/dt$, the  expectation of an arbitrary operator  becomes
  \begin{equation}
 	\braket{\mathcal{O}(t)}=\braket{e^{i B_I(t)}\mathcal{O}_I(t) e^{-i B_I(t)}}\approx 	\braket{O_I(t)}-i\braket{[\mathcal{O}_I,B_I]}+\cdots,
 \end{equation}
 where the subscript $I$ denotes operators in the interaction picture, which we adopt in what follows.  If $\mathcal{O}$ and  $B$ do not commute, the boundary term does impact the expectation at first order in the interaction. In ``conventional" theories, in which the Lagrangian is just a function of the configuration variables and its \emph{first} time derivatives, $L=L(q^i,\dot{q}^i)$ a boundary term can only depend on the  $q^i$, and thus does not contribute to the expectation of any $q$-dependent observable.  On the other hand, the counterterms in the gravitational sector  contain \emph{second} derivatives of the metric, which upon integration by parts result in boundary terms that depend on  $\delta \dot g_{\mu\nu}$.  Since  the latter are generally  proportional to  canonical momenta,  we expect these to give  corrections to the expectation values of $\delta g_{\mu\nu}$.
  
 In this appendix we are mainly concerned with the corrections to the expectation of  $\delta \phi$ and $\delta g_{\mu\nu}$ stemming from the boundary contributions of the counterterms. The latter appear when their variation is  integrated by parts to isolate terms proportional to the undifferentiated fields.  Because the counterterms themselves are proportional to formally divergent constants, it is important to appropriately handle and remove their eventually divergent contributions to the quantum-corrected equations of motion.  As a way of a summary, it turns out that the counterterms we had to introduce do not yield any boundary factors proportional to $\delta\dot \phi$ or $\delta \dot N$. This  is why we have chosen to focus in the main text on the  analogues of the inflaton equation of motion and the Friedman equations, which follow from demanding $\braket{\delta \phi}=\braket{\delta N}$. On the other hand, the counterterms do yield factors at the boundary  proportional to $\delta \dot a$, which cannot be canceled by any boundary  action with the required symmetry.  
 
 \subsection{External Gravitational Field}  
 
 Let us begin with the corrections to the evolution of the inflaton $\phi$ that we studied in Section \ref{sec:Backreaction on the Inflaton Motion}. There we assumed that the gravitational field was fixed, and determined that the divergencies from matter loops could be canceled by the counterterms in equation (\ref{eq:counterterms 1}). Because these do not contain any derivatives of $\phi$, there are no boundary terms to consider.

 \subsection{Dynamical  Gravitational Field}   
 
But things are not as simple when we take the dynamics of the gravitational field into account. In order to determine how  boundary terms impact the expectation  $\delta g_{\mu\nu}$, and thus how the background spacetime evolves, we first need to make assumptions about  the dynamics of the gravitational sector, as dictated by general relativity.  Its action is invariant under diffeomorphisms, so in order to define the propagator we need to fix the gauge. For the purposes of this discussion, it will be simpler to work  in  the ADM formulation \cite{Arnowitt:1962hi}, in which the  field variables are the lapse function $N$, the shift vector $N^i$ and the spatial metric $g_{ij}$.   Which of these becomes dynamical depends on the choice of gauge.

We shall restrict our attention to the tree-level corrections to the field expectation. In that case, it suffices to consider interactions linear in the  field perturbations. Because we are dealing with a  cosmological background, these linear terms are invariant under spatial translations and rotations, so we way restrict our attention to  space-independent scalars under rotations,
\begin{equation}\label{eq:ADM metric}
 	ds^2=-N(t)^2 dt^2 +a^2(t)\delta_{ij} dx^i dx^j, \quad \phi=\phi(t).
\end{equation}
With this metric the Einstein-Hilbert action  action becomes
\begin{equation}\label{eq:S EG GHY}
	S_{EH}+S_{GHY}=\mathcal{V}\int dt \left(-3M_P^2 \frac{a \dot{a}^2}{N}+\frac{a^3 \dot{\phi}^2}{2N}-a^3 N V(\phi)\right),
\end{equation}
to which we  have added  a Gibbons-Hawking-York boundary term \cite{York:1972sj,Gibbons:1976ue} to cancel a boundary contribution from the Einstein-Hilbert action proportional to $\dot{a}$. In this form the theory only contains up to first derivatives of the  variables. In particular,  note that the action (\ref{eq:S EG GHY}) does not contain any time derivatives of $N$, which is auxiliary.  Recall that $\mathcal{V}$ is the volume of the spatial section of our compact universe. 

In order to study the quantum corrected equations of motion, we shall split  the variables  in equation (\ref{eq:ADM metric}) into background plus perturbations,
\begin{equation}
 	N=\bar{N}+\delta N, \quad
	a=\bar{a}+\delta a, \quad
	\phi=\bar\phi+\delta\phi. 
\end{equation}
 Since the action is invariant under a local symmetry,  time diffeomorphisms, we need to fix the gauge (or work in the reduced space of gauge-invariant functions, which we shall avoid.)  Simple  canonical gauges  such as $\delta a=0$   or $\delta\phi=0$   are not  suited for our purposes, because our goal is to calculate  the expectation of $\delta a$ or $\delta\phi$, rather than to fix  it. We shall consider instead  a family of derivative gauges by adding to the action the gauge-breaking term
 \begin{equation}
  	S_{\cancel{\mathrm{gauge}}}=-\mathcal{V}\int dt\,  \frac{(\delta\dot{N}+\chi)^2}{2},
 \end{equation}
 where $\chi$ is an arbitrary gauge-fixing function of the remaining canonical variables, such as $\chi=\delta a$,  $\chi=\delta \phi$ or even $\chi=0$. We omit the associated ghost terms since we shall not need them. The advantage of this choice is that the former auxiliary variable $\delta N$ can now be regarded as an ordinary dynamical variable.  In the Hamiltonian formulation, then, its canonical momentum $\delta b\equiv{-\mathcal{V}(\delta\dot{N}+\chi)}$ commutes with all  the dynamical variables but $\delta N$. 
 
 As in sections  \ref{sec:Backreaction on the Inflaton Motion} and \ref{sec:Backreaction on the Metric}, we shall obtain the quantum corrected equations of motion by demanding 
 \begin{equation}
 \braket{\delta N}=\braket{\delta a}=\braket{\delta\phi}=0.
\end{equation}
We stated earlier that at tree level in the graviton and the inflaton it is  sufficient to consider vertices linear in these fields. Therefore, as long as the gauge-fixing term is quadratic in the  perturbations,  it actually has no impact on our analysis of the quantum-corrected equations of motion.

 \subsubsection{Two Derivatives}
Consider now the Einstein-Hilbert counterterm in equation (\ref{eq:counterterms 2}). By direct substitution of the metric (\ref{eq:ADM metric}) and integration by parts we find a contribution at the boundary proportional to the variation of the trace of the extrinsic curvature,\footnote{The unit normal to $\partial\mathcal{M}$ points inward, $n_\mu=(N,\vec{0})$, and the extrinsic curvature is defined by ${K_{\mu\nu}=(\delta_\mu{}^\rho+n_\mu n^\rho) \nabla_\rho n_\nu}$. Note that the unit normal we use in \cite{Armendariz-Picon:2019csc} has the opposite sign, and so does the extrinsic curvature.} which in our background equals ${K=-3\, \dot{a}/(Na)}$,
\begin{equation}\label{eq:GHY div}
\delta S_\mathrm{ct}\supset - \delta M_P^2 \,  \mathcal{V}  \, (a^3 \, \delta K)|_{\partial\mathcal{M}}.
\end{equation}
 Because the latter contains a factor of $\delta\dot{a}$, there is a tree-level contribution to the expectation $\braket{\delta a}$ proportional to $[\delta a, \delta \pi_a]$.
 Therefore, such a boundary gives a correction to  the expectation proportional to the divergent $\delta M_P^2$.  The appearance of this new divergence demands further renormalization. Inspection of equation (\ref{eq:GHY div}) immediately suggests that these divergencies can be canceled by the addition of a  counterterm proportional to the Gibbons-Hawking-York boundary action
 \begin{equation}\label{eq:GHY}
 	S_\mathrm{ct}\to S_\mathrm{ct}+\delta M_P^2 \int_{\partial\mathcal{M}} \!\!\! d^3 x \sqrt{\gamma}\, K,
 \end{equation}
 where $\gamma$ is the determinant of the spatial metric.  In principle we  could add an arbitrary finite piece to $\delta M_P^2$ in the last expression, but in the absence of a better criterion to fix this finite contribution, it appears reasonable to demand  exact cancellation. To our knowledge, this problem has only been discussed tangentially in the literature, see for instance \cite{Barvinsky:1995dp} for an early discussion of related matters. In  reference \cite{Armendariz-Picon:2019csc} we noted that such a counterterm needed to be added to the action in order to renormalize the power spectrum of gravitational waves.

To further study the kind of boundary counterterms we may need to introduce as we proceed, we note that the action (\ref{eq:action}) is the integral of a spacetime scalar $\mathcal{L}$. Therefore, under infinitesimal diffeomorphisms generated by the vector $\xi^\mu$, it changes by a boundary term
 \begin{equation}
 	\Delta S=\int_\mathcal{M} \!\!\! d^4 x \sqrt{-g}\, \nabla_\mu (  \xi^\mu  \mathcal{L})=
	\int_{\partial \mathcal{M} } \!\!\!  d^3x \sqrt{\gamma} \, n_\mu \,  \xi^\mu \mathcal{L},
 \end{equation}
 where $n_\mu$ is the  normal to the boundary $\partial\mathcal{M}$. Therefore, the action is only  invariant under  diffeomorphisms that vanish at the boundary, $\delta\xi|_{\partial\mathcal{M}}=0$, or those that  map the boundary onto itself, $\xi^\mu n_\mu|_{\partial\mathcal{M}}=0$. We shall refer to the latter  as ``boundary diffeomorphisms." To the extent that invariance under diffeomorphisms constrains the form of the divergencies that may appear in the theory, the previous argument suggests that divergent terms may also include a divergent  action at the boundary. The latter should be invariant under boundary  diffeomorphisms, and should thus consist of tensors  intrinsic to  the boundary  itself or the bulk spacetime. Those intrinsic to the boundary  include the normal $n_\mu$, the spatial metric $\gamma_{\mu\nu}$, the extrinsic curvature $K_{\mu\nu}$ as well as their spatial covariant derivatives. Derivatives of these tensors along the normal are not allowed because these would not be invariant under boundary diffeomorphisms.  Tensors intrinsic to the spacetime include the scalar $\phi$, the spacetime metric and the Riemann tensor $R^\mu{}_{\nu\rho\sigma}$. The boundary term (\ref{eq:GHY}) is precisely of this form. 
 
Let us  consider  now the counterterm proportional to $\delta\xi$ in equation (\ref{eq:counterterms 1}). Integrating by parts to isolate bulk contributions proportional to the undifferentiated field perturbations we arrive at the boundary term
\begin{equation}\label{eq:nm boundary}
\delta S_\mathrm{ct}\supset 2 \delta\xi \, \mathcal{V}\, a^3 \,\lambda(\phi) 
	\, \delta K |_{\partial\mathcal{M}}.
\end{equation}
Such an interaction would again contribute a divergent factor  to $\braket{\delta a}$.  It is clear then that this contribution can be eliminated by the addition of a boundary action to the counterterms,
\begin{equation}
S_\mathrm{ct}\to S_\mathrm{ct}-2\delta \xi  \int_{\partial\mathcal{M}} \!\!\! d^3 x \sqrt{\gamma}\, \lambda(\phi) \, K.\end{equation}
Again, the cancellation does not fix the finite piece of the boundary counterterm, which we set to zero in the absence of   better guidance.

 \subsubsection{Four Derivatives}
 
 At four derivatives the structure of the counterterms is more interesting, albeit somewhat more complicated. 
In order to isolate the terms that contribute to the expectation of $\delta N$ or $\delta a$, it suffices to  focus on terms that contain their  time derivatives, while ignoring those without. 
Thus isolating the boundary contributions of the four-derivative counterterms in equation (\ref{eq:counterterms 2}) we find
 \begin{subequations}\label{eq:bulk variations}
 \begin{align}
 \delta\!\left( \int d^4 x \sqrt{-g} R^2 \right)&\supset 
 	72\, \mathcal{V} \,   
	\left(\frac{\dot{a}^2}{N^2}-\frac{a\dot a\dot N}{N^3}+\frac{a \ddot{a}}{N^2}\right)
	\frac{\delta\dot a}{N} \bigg|_{\partial\mathcal{M}}, \\
 \delta \! \left( \int d^4 x \sqrt{-g} R_{\mu\nu} R^{\mu\nu}\right)& \supset 
  	  12\, \mathcal{V}   \left(
	  \frac{\dot{a}^2}{N^2}-\frac{2a\dot a \dot N}{N^3}+\frac{2a\ddot{a}}{N^2}
	  \right)\frac{\delta\dot a}{N}  \bigg|_{\partial\mathcal{M}}.
 \end{align}
 For further illustration, it shall also be useful to determine the variation of the Riemann squared
  \begin{equation}
  \delta\!\left( \int d^4 x \sqrt{-g}  R^\mu{}_{\nu\rho\sigma}  R_\mu{}^{\nu\rho\sigma} \right)
 \supset  24\, \mathcal{V} \left(
 \frac{a\ddot a}{N^2}-\frac{a\dot{a}\dot{N}}{N^3}
 \right)\frac{\delta \dot a}{N}  \bigg|_{\partial\mathcal{M}}.
  \end{equation}
  \end{subequations}
 These are the terms that we may need to cancel with the appropriate boundary counterterms. Because the bulk variation does not contain any  factors of $\delta\ddot N$, there is no factor proportional to $\delta \dot N$ at the boundary, and we do not need to worry about divergent corrections to $\braket{\delta N}$.  
 
The boundary terms in equations (\ref{eq:bulk variations})  contain a total of three time derivatives of the spatial  metric. Therefore, invariance under boundary diffeomorphisms implies that the boundary action ought to be of the form
 \begin{equation}\label{eq:boundary ct ansatz}
 	S_\mathrm{ct}^{(4)}=\int d^3 x \sqrt{\gamma} \sum_\alpha \delta b_\alpha L_\alpha,
 \end{equation}
 with  the $L_\alpha$ chosen from the set of all spatial scalars with three time derivatives acting on $g_{\mu\nu}$. Because the Riemann tensor components can be expressed in terms of those of the extrinsic curvature, such a set can be chosen to consist, without loss of generality, of the five invariants
 \begin{equation}\label{eq:b ct set}
 	1. \, K^3, \quad
	2.\, K_{ij} K^{ij} K,  \quad
	3.\,  K_i{}^j K_j{}^k K_k{}^i, \quad
	4.\,  n^\mu n^\nu R_{\mu\nu} K,  \quad
 	5.\, n^\mu n^\nu R_{i\mu j\nu} K^{ij}.
\end{equation}
 A term $n^\mu \nabla_\mu R$ would be allowed too, but it cannot belong to the set of  counterterms because its variation would contain a boundary factor proportional to $\delta\ddot{a}$, which has no counterpart in equations (\ref{eq:bulk variations}). Similarly,  there is no need to consider the Riemann tensor of the spatial boundary, since it does not contain any time derivatives, and it vanishes in our background anyway. 

In order to find  the coefficients $\delta b_\alpha$ we simply need to match equations (\ref{eq:bulk variations}) to the variation  of equation (\ref{eq:boundary ct ansatz}), keeping in mind that we only need to focus on terms that contain derivatives of $\delta N$ or $\delta a$. Note that in our symmetric background many of the terms in equation (\ref{eq:boundary ct ansatz}) are proportional to each other, since
\begin{equation}
	K_{ij}=-\frac{a\dot a}{N}\delta_{ij}, 
	\quad
	n^\mu n^\nu R_{\mu\nu}=\frac{3}{a N^2}\left(\frac{\dot a \dot N}{N}-\ddot{a}\right),
	\quad
	n^\mu n^\nu R_{\mu i \nu j}=\left(\frac{\dot a \dot N}{N}-\ddot{a}\right)\frac{a\, \delta_{ij}}{N^2}.
\end{equation}
This implies that we can set, without loss of generality $\delta b_2=\delta b_3=\delta b_5=0$, and we only need to contend with two independent boundary terms, $K^3$ and $ n^\mu n^\nu R_{\mu\nu} K$. Because  equations (\ref{eq:bulk variations}) do not contain any derivatives of $\delta N$, it is then easy to see that $\delta b_4$ must vanish, which implies that $L_4$  does not contribute to the variation in (\ref{eq:boundary ct ansatz}). Therefore, the only remaining term in the latter is
\begin{equation}
\delta S_{\mathrm{ct}}^{(4)}=-81 \mathcal{V} \, \delta b_1 \frac{\dot{a}^2}{N^2}\frac{\delta\dot a}{N}
	 \bigg|_{\partial\mathcal{M}},
\end{equation}
which needs to equal the boundary contribution  from the counterterms in equations (\ref{eq:bulk variations}). This is only possible if
\begin{subequations}
\begin{align}
	24\delta c_{(1)}+4\delta c_{(2)}&=27\delta b_1,
	\\
	3\delta c_{(1)}+\delta c_{(2)}+\delta c_{(3)}&=0.  \label{eq:boundary c1 and c2}
\end{align}
\end{subequations}

Unfortunately, equation  (\ref{eq:boundary c1 and c2}) is  incompatible with the structure of the radiative corrections that we have studied. Namely, in section \ref{sec:Evaluation of the Energy Density} we have seen  that one can renormalize the bulk divergencies  of the theory by setting $\delta c_{(3)}=0$ if $\delta c_{(1)}$ and $\delta c_{(2)}$ obey equation (\ref{eq:delta c1 and c2}).  But these conditions are incompatible with (\ref{eq:boundary c1 and c2}).  More generally, equation (\ref{eq:boundary c1 and c2})  is also in conflict with with the structure of the counterterms known from the one-loop effective action of pure gravity \cite{tHooft:1974toh}, which can be taken to be ${\delta c_{(2)}=2\delta c_{(1)}}$, $\delta c_{(3)}=0$,

A set of values that does obey equation (\ref{eq:boundary c1 and c2}) is  $\delta c_{(2)}=-4\delta c_{(3)}=-4\delta c_{(1)}$, which correspond  to the Gauss-Bonnet action 
\begin{equation}
S_{GB}=\int d^4 x\sqrt{-g}\left[R^2-4R_{\mu\nu} R^{\mu\nu}+R_{\mu\nu\rho\sigma}R^{\mu\nu\rho\sigma}\right].
\end{equation}
Its variation is known to be a boundary term, which  can  indeed be cast as the variation of equation (\ref{eq:boundary ct ansatz}) when $\delta g_{\mu\nu}=0$ but $\delta\dot g_{\mu\nu}\neq 0$ \cite{Deruelle:2017xel}. The symmetry of our background prevents us from reaching further conclusions, but it is likely that the Gauss-Bonnet action is the only  combination of quadratic  curvature invariants with such property.

\subsubsection{Relation to the Variational Principle}

Our quandaries are related to the presence of higher derivative terms in the action, and they are also connected to the eventual absence of a well-posed variational principle in these theories \cite{Dyer:2008hb}. Recall that a well-posed variational principle  implies the existence of an extremum of the action  when the variation of the variables $\delta q$ is constrained to vanish at the boundary. Since the variation of $\delta\dot q$ is left unconstrained, it therefore cannot appear in a boundary term.  In such theories, the demand $\braket{\delta q}=0$ leads to the classical equations of motion, because the boundary term does not contribute to the expectation of $\braket{\delta q}$, as we discussed above. If the theory contains higher derivatives, however, terms proportional to $\delta \dot q$ may appear  at the boundary. These theories do not have a well-posed variational principle. In those cases, we also expect a contribution from the boundary terms to the expectation of $\delta q$, because $[\delta q, \delta \dot{q}]$ is  typically non-zero. In other words, what we have noted here is that, when taken at face value, a theory of the from  given by (\ref{eq:counterterms 2})  does not admit a well-posed variational principle, even if we allow for the addition of appropriate boundary terms.  In some particular cases, such as $\delta c_2=\delta c_3=0$, this has been previously noted in the literature \cite{Madsen:1989rz}.
\end{fmffile}

\end{document}